\documentclass[twocolumn]{aa}

\usepackage{graphicx}
\usepackage{txfonts}
\usepackage[usenames,dvipsnames,svgnames,table]{xcolor}
\usepackage{booktabs}
\usepackage{multirow}
\usepackage{amssymb}
\usepackage{enumitem}
\usepackage{hyperref}
\usepackage[all]{hypcap}
\hypersetup{
    colorlinks,
    linkcolor={red!50!black},
    citecolor={blue!50!black},
    urlcolor={blue!80!black}
}

\begin{document}

\title{The ionization fraction in OMC-2 and OMC-3} 

\titlerunning{The ionization fraction in OMC-2 and OMC-3}

\author{P.~Salas\inst{1}\fnmsep\thanks{psalas@nrao.edu}
        \and
        M.~R.~Rugel\inst{2}
        \and
        K.~L.~Emig\inst{3}
        \and
        J.~Kauffmann\inst{4}
        \and
        K.~M.~Menten\inst{2}
        \and
        F.~Wyrowski\inst{2}
        \and
        A.~G.~G.~M.~Tielens\inst{5}
        }

\institute{Green Bank Observatory, 155 Observatory Road, Green Bank, WV 24915, USA
            \and
            Max-Planck-Institut fur Radioastronomie, Auf dem H{\"u}gel 69, D-53121 Bonn, Germany
            \and
            National Radio Astronomy Observatory, 520 Edgemont Road, Charlottesville, VA 22903, USA
            \and
            Haystack Observatory, Massachusetts Institute of Technology, Westford, MA 01886, USA
            \and
            Leiden Observatory, Leiden University, P.O. Box 9513, NL-2300 RA Leiden, The Netherlands
            }

\date{\today}

\abstract {The electron density ($n_{\mathrm{e}^{-}}$) plays an important role in setting the chemistry and physics of the interstellar medium.
However, measurements of $n_{\mathrm{e}^{-}}$ in neutral clouds have been directly obtained only toward a few lines of sight or they rely on indirect determinations.} 
{We use carbon radio recombination lines and the far-infrared lines of C$^{+}$ to directly measure $n_{\mathrm{e}^{-}}$ and the gas temperature in the envelope of the integral shaped filament (ISF) in the Orion A molecular cloud.}
{We observed the C$102\alpha$ ($6109.901$~MHz) and C$109\alpha$ ($5011.420$~MHz) carbon radio recombination lines (CRRLs) using the Effelsberg 100m telescope at ${\approx}2\arcmin$ resolution toward five positions in OMC-2 and OMC-3.
Since the CRRLs have similar line properties, we averaged them to increase the signal-to-noise ratio of the spectra.
We compared the intensities of the averaged CRRLs, and the $158~\mu$m-[CII] and [$^{13}$CII] lines to the predictions of a homogeneous model for the C$^{+}$/C interface in the envelope of a molecular cloud and from this comparison we determined the electron density, temperature and C$^{+}$ column density of the gas.}
{We detect the CRRLs toward four positions, where their velocity ($\varv_{\rm{LSR}}{\approx}11$~km~s$^{-1}$) and widths ($\sigma_{\varv}{\approx}1$~km~s$^{-1}$) confirms that they trace the envelope of the ISF.
Toward two positions we detect the CRRLs, and the $158~\mu$m-[CII] and [$^{13}$CII] lines with a signal-to-noise ratio $\geq5$, and we find $n_{\mathrm{e}^{-}}{=}0.65\pm0.12$~cm$^{-3}$ and $0.95\pm0.02$~cm$^{-3}$, which corresponds to a gas density $n_{\mathrm{H}}{\approx}5\times10^{3}$~cm$^{-3}$ and a thermal pressure of $p_{\mathrm{th}}{\approx}4\times10^{5}$~K~cm$^{-3}$.
We also constrained the ionization fraction in the denser portions of the molecular cloud using the HCN$(1\mbox{--}0)$ and C$_{2}$H$(1\mbox{--}0)$ lines to $x(\mathrm{e}^{-}){\leq}3\times10^{-6}$.}
{The derived electron densities and ionization fraction imply that $x(\mathrm{e}^{-})$ drops by a factor ${\geq}100$ between the C$^{+}$ layer and the regions probed by HCN$(1\mbox{--}0)$.
This suggests that electron collisional excitation does not play a significant role in setting the excitation of HCN$(1\mbox{--}0)$ toward the region studied, as it is responsible for only ${\approx}10\%$ of the observed emission.}

\maketitle

\section{Introduction}

The electron density, $n_{\mathrm{e^{-}}}$, plays a fundamental role in the physics, chemistry, and dynamics of the neutral (atomic or molecular) interstellar medium (ISM). Collisions with electrons ejected through the photo-ionization of polyciclic aromatic hydrocarbons (PAHs) and small dust grains provide the main heating mechanism in regions exposed to far-ultraviolet (FUV) photons with energies $<13.6$~eV \citep[e.g.,][]{Watson1972,Hollenbach1999}. The electron density also determines the rate of fast ion-neutral reactions in molecular clouds, one of the main channels for the build-up of molecular complexity \citep[e.g.,][]{Herbst1973,Oppenheimer1974}. In addition, the abundance of electrons relative to the neutrals, the ionization fraction  ($x(\mathrm{e^{-}})=n_{\mathrm{e^{-}}}/n_{\mathrm{H}}$), mediates the coupling between neutral gas and magnetic fields through ion-neutral friction \citep[e.g.,][]{Mestel1956}. Finally, large dipole moment molecules, such as HCN and CN, have large inelastic collision cross sections with electrons, which leads to the significant rotational excitation of these molecules in regions where $x(\mathrm{e^{-}})\geq10^{-5}$ \citep[e.g.,][]{Dickinson1977,Black1991,Goldsmith2017}.

The regions of the neutral ISM where FUV photons regulate the heating and chemistry are known as photodissociation regions (PDRs). In the outer layers of PDRs, the gas is neutral and carbon is singly ionized since it has a lower ionization potential than hydrogen ($11.2$~eV). This makes carbon one of the main donors of free electrons in the outer layers of PDRs. As we move deeper into a PDR, away from the source of FUV photons, the number of FUV photons drops and C$^{+}$ recombines with free electrons in the C$^{+}$/C interface. After recombination, carbon can be found in a highly excited electronic state as a Rydberg atom, and while the electron cascades to the ground energy levels it will emit at radio frequencies. The lines observed from this cascade are known as carbon radio recombination lines \citep[CRRLs, ][]{Gordon2009}, and along with the far-infrared line of C$^{+}$ at $158~\mu$m, [CII], they are important tracers of PDRs. Since the intensities of [CII] and CRRLs have different dependencies on the gas temperature and density, their combined use is a powerful probe of the gas properties in a PDR \citep[e.g.,][]{Natta1994,Sorochenko2010,Tsivilev2014,Salgado2017b,Cuadrado2019,Salas2019,Siddiqui2020}, such as the envelopes of molecular clouds.

\begin{figure*}[ht]
  \resizebox{\hsize}{!}
  {\includegraphics[width=0.45\textwidth]{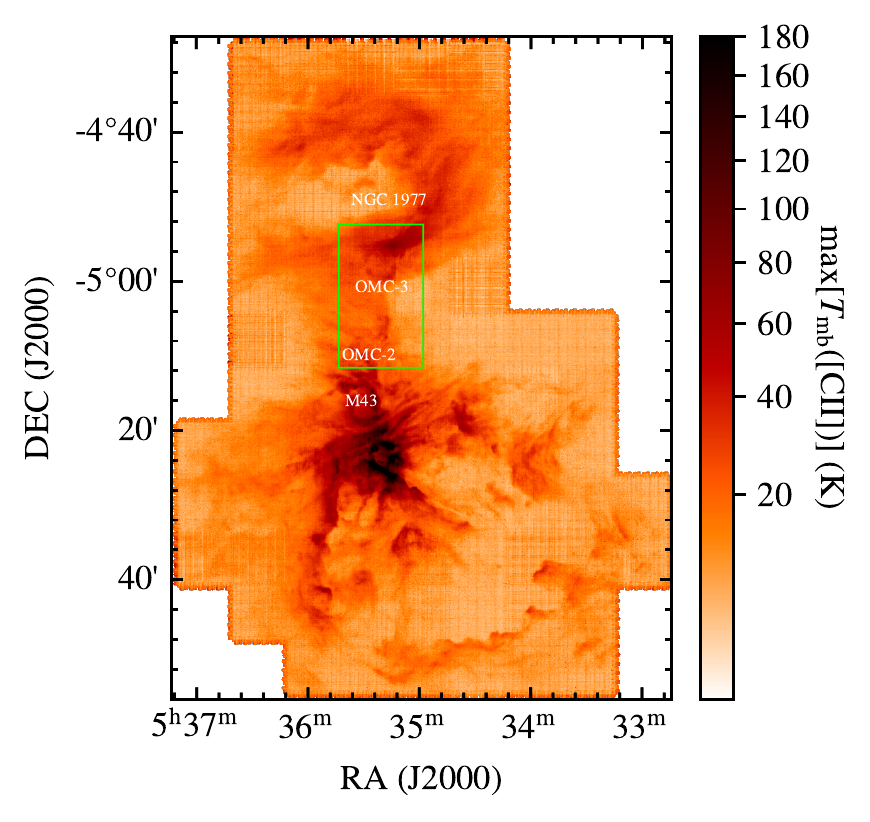}
   \includegraphics[width=0.3\textwidth]{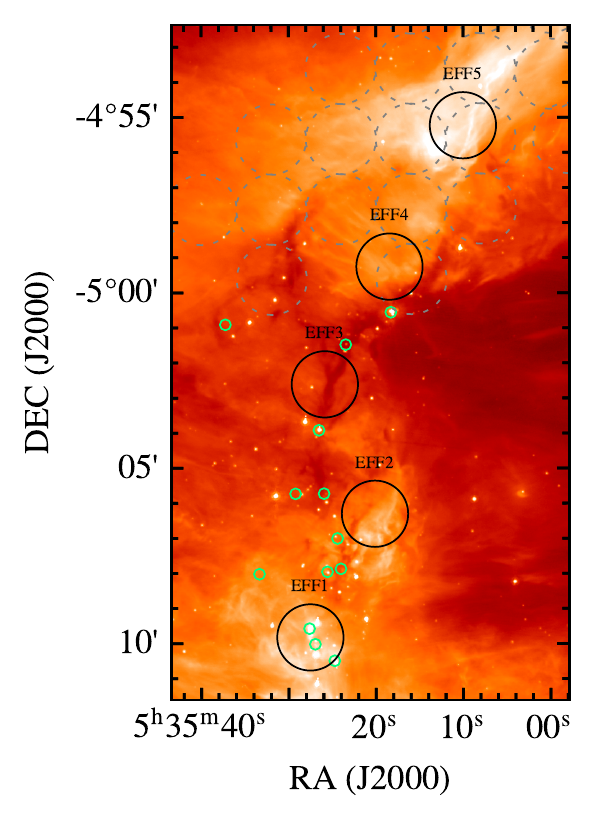}} %
  \caption{Location of the Effelsberg $5$~GHz RRL observations presented in this work.
  Left: Map of the peak temperature of the [CII] line over a one square degree centered on the Trapezium stars \citep{Pabst2019}.
  A green rectangle shows the extent of the region displayed on the right.
  Right: Spitzer IRAC $8~\mu$m image of the OMC-2/3.
  The regions observed in $5$~GHz RRLs are marked with solid black circles $1\farcm89$ in diameter, while the dashed gray circles show the C$76\alpha$ observations of \citet{Kutner1985} at $2\arcmin$ resolution.
  The small green circles show the location of the sources detected in $3.6$~cm radio continuum \citep{Reipurth1999}.}
  \label{fig:beams}
\end{figure*}

Deeper into the PDR, past the C$^{+}$/C interface, most of the carbon is in the form of CO and the electron density is mainly set by cosmic-ray ionization. In the dense cores of molecular clouds, the cosmic ray ionization rate (CRIR) is typically ${\sim}10^{-17}$~s$^{-1}$ \citep[][]{Caselli1998,vanderTak2000}, leading to $x(\mathrm{e^{-}}){\sim}10^{-9}$--$10^{-6}$ \citep[e.g.,][]{Caselli1998,Goicoechea2009}. In a stationary, onion-like, view of molecular clouds it is in these dense cores that the abundances of molecules are high enough for their emission to be detected.
However, porosity of the PDR to FUV radiation might leave the cold clouds exposed to FUV radiation. Measuring the ionization fraction at different depths in a PDR provides insight into the structure of PDRs, the penetration of FUV radiation and the excitation of molecules with large dipole moments, for example, HCN.

Due to its high optically thin critical density \citep[${\sim}10^{5}$~cm$^{-3}$, e.g.,][]{Shirley2015}, HCN$(1\mbox{--}0)$ has been used as a tracer of dense molecular gas \citep[$n_{\mathrm{H}_2}\gtrsim3\times10^{4}$~cm$^{-3}$, e.g.,][]{Gao2004}. However, recent studies of molecular clouds in our Galaxy show that a large fraction of the HCN$(1\mbox{--}0)$ emission traces regions where the gas density is ${\sim}10^{3}$~cm$^{-3}$ \citep[e.g.,][]{Kauffmann2017,Pety2017}. Since the gas density is an essential parameter in our theories of star formation \citep[e.g.,][]{McKee2007}, it is important to determine the characteristic density of the gas traced by HCN, and what processes dominate the excitation of HCN$(1\mbox{--}0)$ at different densities. Some of the mechanisms responsible for the HCN$(1\mbox{--}0)$ emission at low densities include additional collisional excitation of the transition due to collisions with electrons in high ionization fraction regions \citep[e.g.,][]{Dickinson1977,Black1991,Goldsmith2017}, an elevated gas temperature due to cosmic ray heating \citep[e.g.,][]{Meijerink2011,Vollmer2017} and radiative trapping due to a large line optical depth \citep[e.g.,][]{Evans1999,Shirley2015}. To assess the importance of these mechanisms it is necessary to obtain independent constraints on the gas properties, such as its temperature and density.

At a distance of $414$~pc \citep[][]{Menten2007,Zari2017} Orion~A is the closest molecular cloud where massive stars have recently formed. In the northern most portion of Orion~A, \citet{Bally1987} identified an integral shaped filament (ISF) ${\sim}13$~pc long. The northern half of the ISF is composed of OMC-2 and OMC-3 \citep[][]{Gatley1974,Morris1974}. Their masses, as inferred from C$^{18}$O observations, are $113$~M$_{\odot}$ and $140$~M$_{\odot}$, respectively \citep{Dutrey1993}, and they cover an area ${\approx}2.5\times2.5$~pc$^{2}$. The mild radiation field and the presence of embedded stars \citep[e.g.,][]{Megeath2012} make OMC-2 and OMC-3 representative of an average star forming molecular cloud in the Milky Way. It is also toward these clouds that \citet{Kauffmann2017} studied HCN$(1\mbox{--}0)$ emission, and by modeling the ISF as a cylinder with a diameter of $2$~pc they conclude that $50\%$ of the HCN$(1\mbox{--}0)$ emission traces regions with densities below ${\sim}10^{3}$~cm$^{-3}$.

In this work we target five regions toward OMC-2/3 to study how the ionization fraction changes from the envelope of the molecular cloud to its denser core. We use CRRLs and [CII] to measure the temperature and electron density of the molecular cloud's envelope, and we use observations of molecular gas tracers (i.e., HCN and C$_{2}$H) to constrain the ionization fractions deeper into the cloud. We use these constraints on $x(\mathrm{e^{-}})$ to determine the role of collisions with electrons in the excitation of the HCN$(1\mbox{--}0)$ line. When quoting uncertainties we give the $1\sigma$ ranges unless otherwise noted.

\section{Observations}

\subsection{$5$~GHz RRLs}

We observed the $n=110$, $109$ and $102$ $\alpha$\footnote{Transitions involving a change in principal quantum number $\Delta n=1$.} CRRLs at $4876.588$, $5011.420$ and $6109.901$~MHz, respectively, with the Effelsberg 100m radio telescope during $2019$ December and $2020$ March. We used the wide-band Effelsberg C-band receiver (S45mm) at the secondary focus, which uses linear polarization feeds. The spectra were recorded with fast Fourier transform spectrometer backends \citep{KleinHochgurtel:2012aa}. They provide two 200~MHz-wide bands with 65536 channels each and a channel width of 38.1~kHz. We place these at 4941.0~MHz to capture the $n=110$ and $109$ $\alpha$ CRRLs, and at 6111.0~MHz for the $n=102\alpha$ CRRL. A summary of the radio recombination lines (RRLs) covered by the spectral setup is presented in Table~\ref{tab:effspec}.

We targeted five regions, shown in Figure~\ref{fig:beams} and presented in Table~\ref{tab:effobs}. The observations were done in position switching, with the reference position listed as ``OFF'' in Table~\ref{tab:effobs}. As flux calibrator we observed NGC 7027 \citep[2105+420,][]{Ott1994} at the beginning of each session. The pointing was checked regularly on 3C161 or 0420-014, with a typical pointing accuracy of $\lesssim$5\arcsec. We also observed M42 as a reference before each observation. 

The observations were reduced with the Effelsberg reduction pipeline, based on standard single-dish techniques as described in \citet{Kraus:2009aa}. We reduced each 200-MHz-wide band separately, and verified that frequency-dependent effects of the bandpass, system temperature and the noise diode \citep{WinkelKraus:2012ab} were marginal ($\lesssim 3\%$) in our setup. The pipeline used the new water vapor radiometer for atmospheric opacity correction \citep{Roy2004}. The native channel width is $0.15$~km~s$^{-1}$ and $0.18$~km~s$^{-1}$ at 6111\,MHz and 4941\,MHz, respectively. Due to gridding of the spectra to the Kinematic Local Standard of Rest (LSRK) frame \citep{Reid2009} in the reduction process the final spectral resolution is degraded by a factor of two with respect to the native channel width. The absolute flux calibration was done by comparing pointing observations with a flux density model of NGC 7027 which is based on regular Effelsberg monitoring (Kraus et al., private comm.). Typical system temperatures during the observations were ${\approx}50$~K. The beam sizes are 114\arcsec\ and 138\arcsec, the sensitivity in K/Jy is 1.63 and 1.66 and the main beam efficiency is $0.65$ and $0.64$ for the bands at $6111$~MHz and $4941$~MHz, respectively, which were used to convert the observed antenna temperatures to main beam temperatures. The absolute flux calibration is accurate to $10$--$20$\%.

\begin{figure*}[h]
  \resizebox{\hsize}{!}
  {\includegraphics[width=0.33\textwidth]{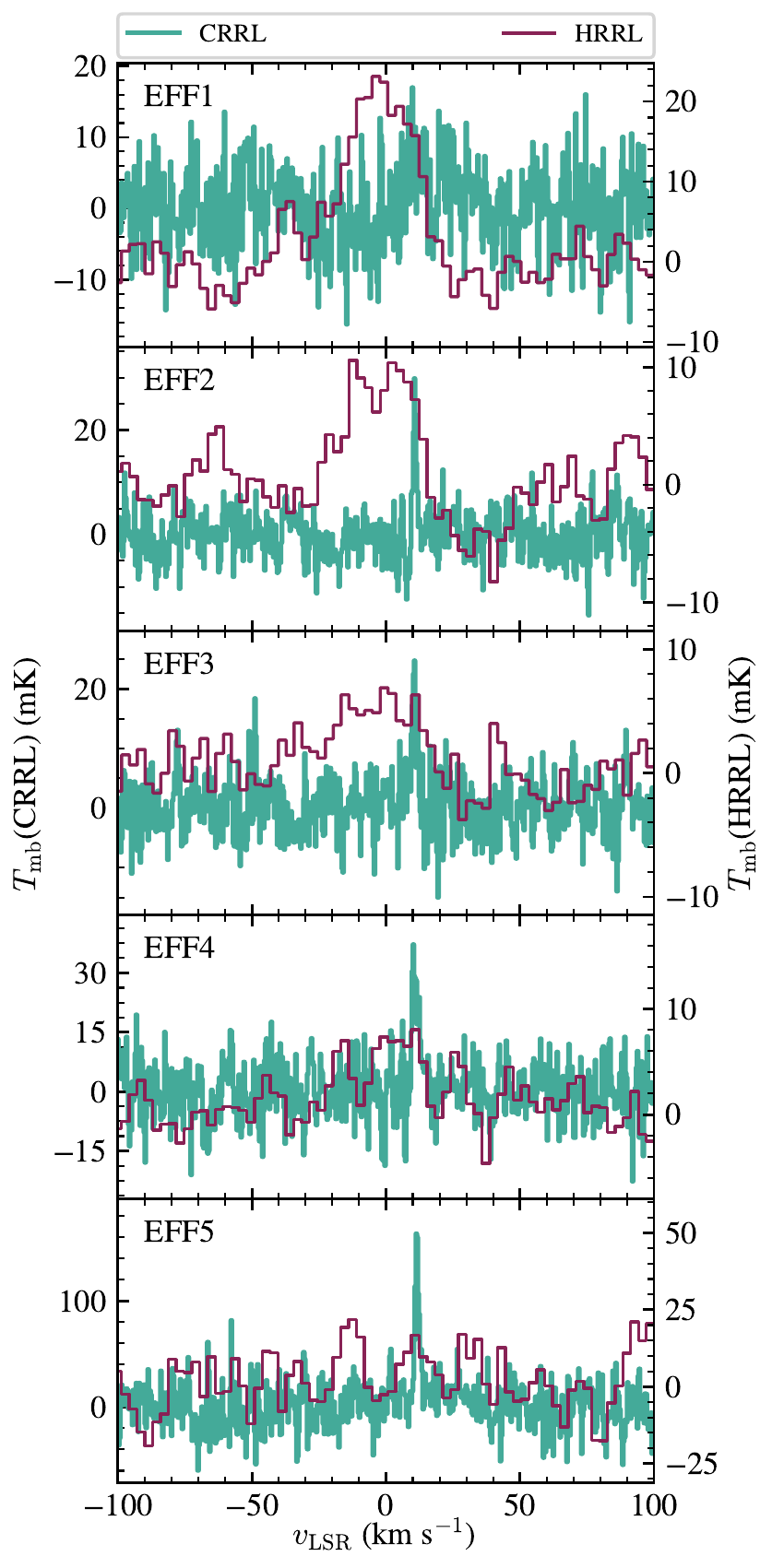}
   \includegraphics[width=0.33\textwidth]{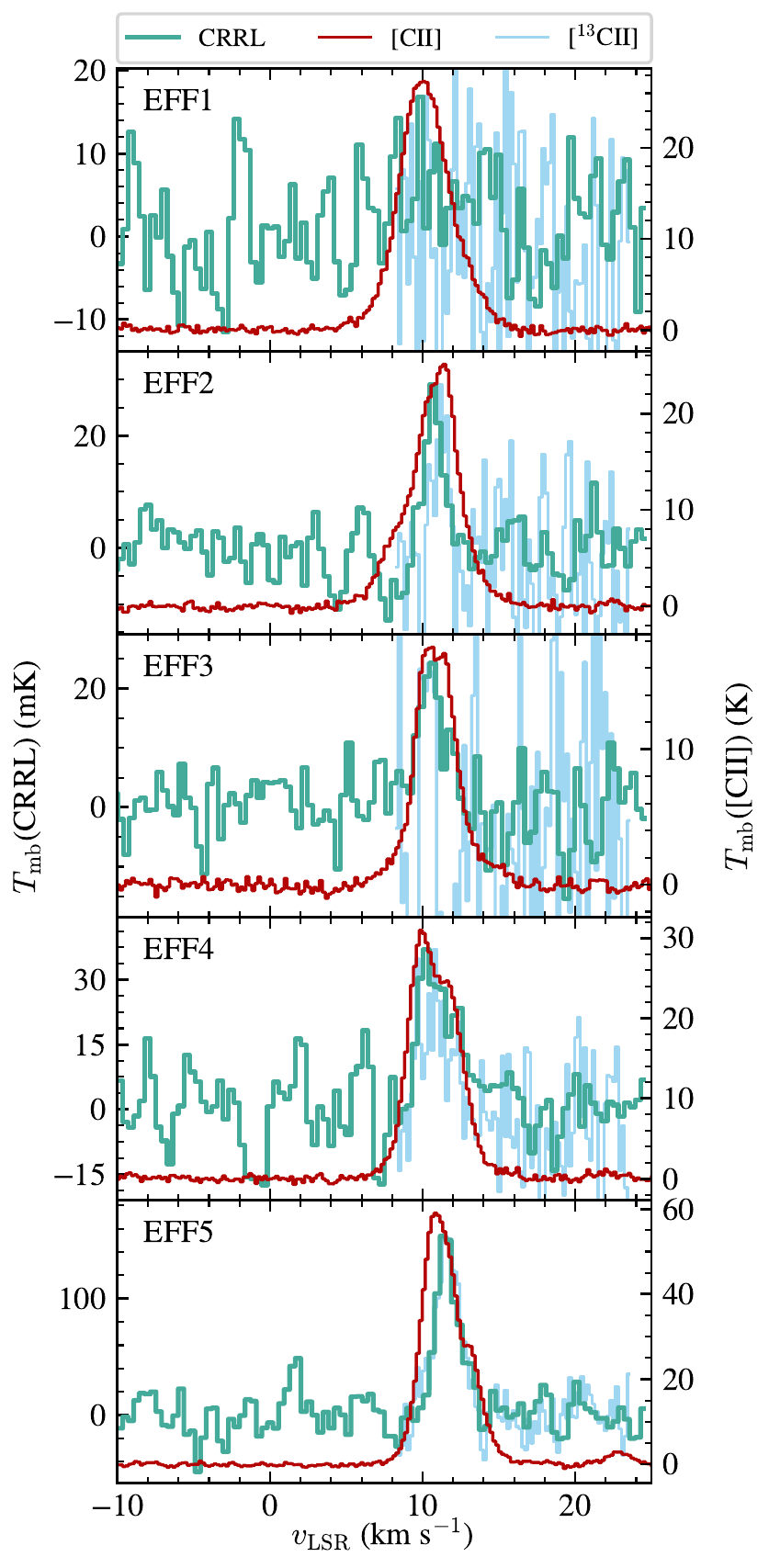}
   \includegraphics[width=0.33\textwidth]{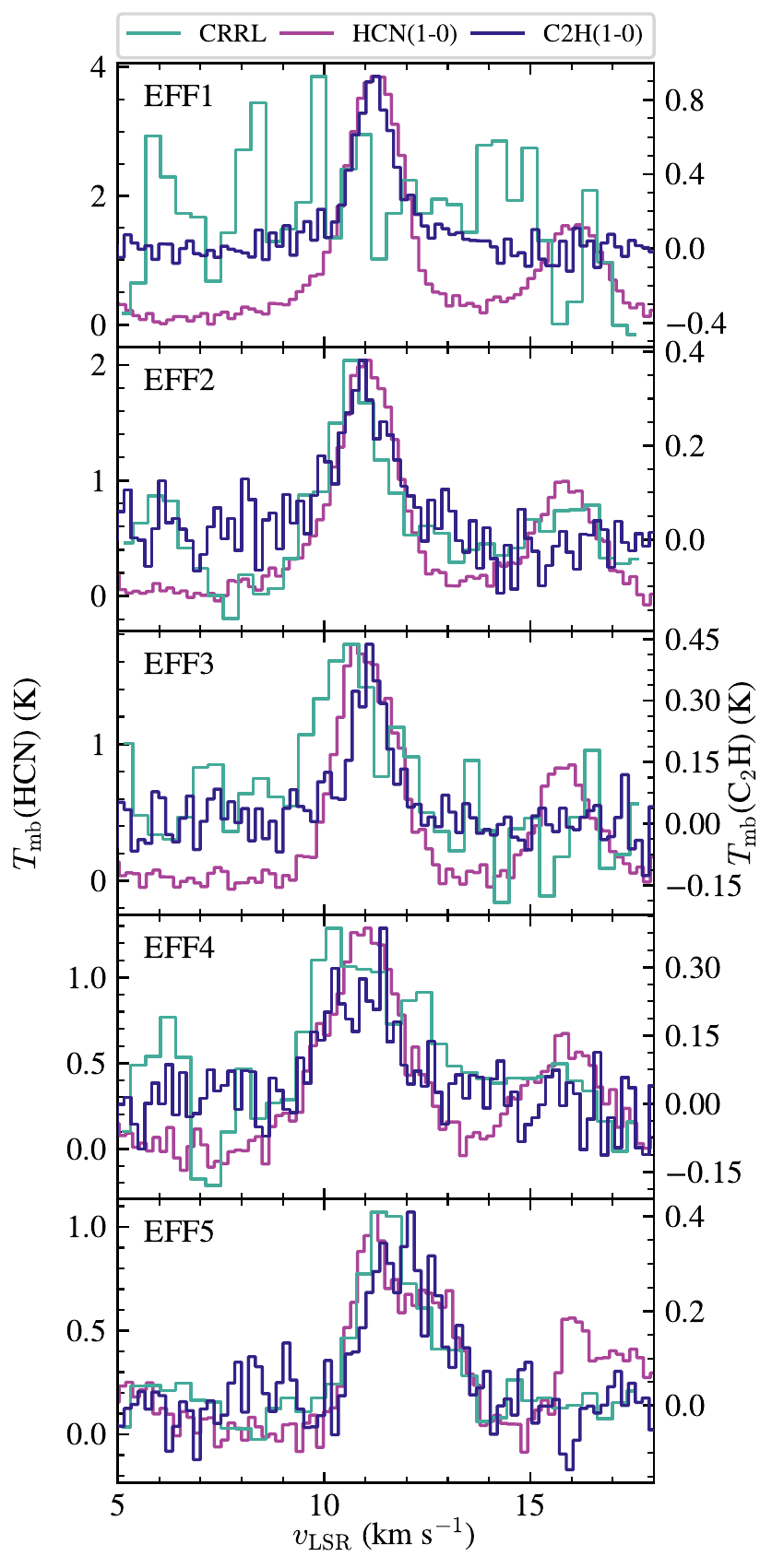}} %
  \caption{CRRL spectra observed toward the northern part of the ISF with the Effelsberg 100m radio telescope.
  The CRRL is an average of the C$109\alpha$ and C$102\alpha$ lines.
  We compare the averaged CRRL to the averaged HRRL (an average of the H$109\alpha$ and H$102\alpha$ lines, on the left), the FIR [CII] and [$^{13}$CII] $F{=}2\mbox{--}1$ lines (center), and HCN$(1\mbox{--}0)$ and C$_{2}$H$(1\mbox{--}0)$ (right).
  In the center panels the [$^{13}$CII] $F{=}2\mbox{--}1$ line is scaled to match the peak of the averaged CRRL, while on the right panels the CRRL is scaled to match the peak of the HCN$(1\mbox{--}0)$ line.}
  \label{fig:isfspectra}
\end{figure*}

Once the spectra have been calibrated we average the C$109\alpha$ and C$102\alpha$ lines (rest frequencies $5011.422$ and $6109.902$~MHz, respectively\footnote{\url{https://splatalogue.online}}), after bringing them to the same velocity grid with a channel size of $0.18$~km~s$^{-1}$ and a velocity resolution of $\sim0.36$~km~s$^{-1}$. The C$110\alpha$ line is contaminated by radio frequency interference, so we do not include it. The averaged spectra have a mean frequency of $5557.9$~MHz, which lies between that of the C$106\alpha$ and C$105\alpha$ lines. In the following we adopt a frequency of $5446.976$~MHz, which corresponds to a transition with a principal quantum number $n=106$, and a beam size of $126\arcsec$ for the averaged CRRL spectra.

The data contain additional ripples at very low intensity ($<0.50$~mK peak-to-trough variations). The ripples are of sinusoidal shape with typical frequencies of $5.8$~MHz (${\sim}300$~km~s$^{-1}$ at the observed frequencies). Since they are much broader than then narrow CRRLs \citep[${\sim}1$~km~s$^{-1}$, e.g.,][]{Cuadrado2019}, we remove a 1st order polynomial baseline around the expected velocity of the CRRLs. To recover information on the detection of hydrogen radio recombination lines (HRRLs), which are broader \citep[${\sim}20$--$30$~km~s$^{-1}$, e.g.,][]{Luisi2019}, we model the baseline on a scan-by-scan basis by fitting a sinus function over at least three periods. Further baselines are removed after stacking. We compare this to interpolating the $5.8$ MHz frequency of the baseline in Fourier space. The HRRLs toward EFF1 and EFF2 are detected in both methods. EFF4 is detected marginally after smoothing and stacking with a signal-to-noise ratio (S/N) of $5$. We do not report detections for EFF3 and EFF5, since EFF3 is not detected when removing the baseline in Fourier space, and EFF5 is detected in neither method.

\begin{table}%[ht]
\caption{\label{tab:effobs} Regions observed with the Effelsberg 100m radio telescope.}
\centering
\begin{tabular}{lccc}
\toprule
Region & RA\tablefootmark{$\dagger$} & DEC\tablefootmark{$\dagger$} & $t_{\rm{int}}$\tablefootmark{\ddag} \\
       & (h:m:s) & (d:m:s) & (minutes) \\
\midrule
OFF    & 5:33:20.00 & -5:05:14.2 & -- \\
EFF1   & 5:35:27.54 & -5:09:49.5 & $156$ \\
EFF2   & 5:35:20.13 & -5:06:18.0 & $245$ \\
EFF3   & 5:35:25.86 & -5:02:36.6 & $217$ \\
EFF4   & 5:35:18.45 & -4:59:15.3 & $118$ \\
EFF5   & 5:35:10.03 & -4:55:13.6 & $12$ \\
\bottomrule
\end{tabular}
\tablefoot{
\tablefoottext{$\dagger$}{J2000 equinox.}\\
\tablefoottext{\ddag}{ON source integration time.}
}
\end{table}

\begin{table}%[ht]
\caption{\label{tab:effspec} Spectral setup of the C-band observations with the Effelsberg 100m radio telescope.\tablefootmark{$\dagger$}}
\centering
\begin{tabular}{lccc}
\toprule
Spectral & Frequency & RRLs\tablefootmark{\ddag}\\
window   & (MHz)     & \\
\midrule
1                  & 6111.0 \tablefootmark{$\dagger\dagger$} & C$102\alpha$, He$102\alpha$, H$102\alpha$ \\
\multirow{2}{*}{2} & \multirow{2}{*}{4941.0}                 & C$109\alpha$, He$109\alpha$, H$109\alpha$ \\
                   &                                         & C$110\alpha$, He$110\alpha$, H$110\alpha$ \\
\bottomrule
\end{tabular}
\tablefoot{
\tablefoottext{$\dagger$}{Both spectral windows are $200$~MHz wide and have $65536$ channels.}\\
\tablefoottext{\ddag}{$\alpha$ RRLs present in each spectral window.}\\
\tablefoottext{$\dagger\dagger$}{On December 11, 2019, the band center was slightly shifted to $6109.0$ MHz}.
}
\end{table}

\subsection{Literature data}

\subsubsection{FIR C$^{+}$}

We use the $158~\mu$m-[CII] line cube presented by \citet{Pabst2019}. This data cube resulted from observations with the upGREAT receiver \citep[][]{Heyminck2012,Risacher2016} onboard the Stratospheric Observatory for Infrared Astronomy \citep[SOFIA][]{Young2012}. The [CII] cube has a spatial resolution of $16\arcsec$ and a velocity resolution of $0.2$~km~s$^{-1}$, and covers a region of roughly $1\degr\times1\degr$. To compare the FIR cube to the radio observations we convolve it to a spatial resolution of 126\arcsec.

\subsubsection{Dust}

To determine the dust conditions toward the regions observed in RRLs we use the temperature and opacity maps of \citet{Lombardi2014}. These maps constructed by combining FIR data obtained with Herschel and Planck and provide the dust properties at a resolution of $36\arcsec$. We use the dust opacity as a proxy for the column density of the neutral gas, $N(\mbox{H})=N(\mbox{H}_{2}+\mbox{HI})$. To convert from optical depth to extinction at K band we adopt their conversion $A_{K}=2640\tau_{850}$~mag, a ratio $A_{K}/A_{V}=0.112$ and $N(\mbox{H})/A_{V}=2.07\times10^{21}$~cm$^{-2}$~mag$^{-1}$. Additionally, we use the dust temperature to estimate $G_{0}$, the FUV radiation field at the surface of the PDR in Habing units \citep[$1.6\times10^{-3}$~erg~s$^{-1}$~cm$^{-2}$,][]{Habing1968}, from $T_{\rm{d}}=12.2G_{0}^{0.2}$~K \citep{Hollenbach1991}.

\subsubsection{Molecular gas}

As a measure of the column density of molecular hydrogen we use the $N(\mbox{H}_{2})$ map of \citet{Berne2014} derived from the $^{12}$CO$(2\mbox{--}1)$ and $^{13}$CO$(2\mbox{--}1)$ lines at ${\approx}11\arcsec$. For the temperature and line widths of the dense molecular gas we use the NH$_{3}$ data cubes produced as part of the Green Bank Ammonia Survey (GAS) data release 1 \citep[][]{Friesen2017}. We rederive the line properties and gas temperature at the spatial resolution of the CRRL observations using \emph{pySpecKit} \citep{Ginsburg2011}. We also use the HCN$(1\mbox{--}0)$ and C$_{2}$H$(1\mbox{--}0)$ $(J{=}1/2\mbox{--}1/2)$ data cubes of \citet{Melnick2011}, as their ratio can be used as a probe of $x(\mathrm{e^{-}})$ \citep{Bron2021}. The C$_{2}$H$(1\mbox{--}0)$ data cube includes the fine structure components $F{=}0\mbox{--}1$ and $F{=}1\mbox{--}1$ within the observed range of frequencies. When comparing the NH$_{3}$, HCN$(1\mbox{--}0)$ and C$_{2}$H$(1\mbox{--}0)$ $(J{=}1/2\mbox{--}1/2)$ lines to the CRRLs we smooth their data cubes to a spatial resolution of $126\arcsec$.

\section{Results}

Here we present the results from our RRL observations, and a comparison between the RRL, [CII] and molecular line spectra.

\subsection{RRLs}

\begin{table*}%[ht]
\caption{\label{tab:c106a} Best fit CRRL properties.}
\centering
\begin{tabular}{lcccccccccc}
\toprule
Region &  $T_{\mathrm{mb}}$ &  $\varv_{\mathrm{c}}$ &  $\Delta\varv$\tablefootmark{a}  & $\int T_{\mathrm{mb}}^{*}dv$ & rms  & S/N\tablefootmark{b} \\
       & (mK)           & (km s$^{-1}$)     & (km s$^{-1}$)& (mK km s$^{-1}$)                                    & (mK) &                      \\
\midrule
  EFF1                   & $<18$      &                &               &               & $6$   &      \\
  \hline
  EFF2                   & $27\pm4$   & $10.74\pm0.08$ & $1.36\pm0.19$ & $40\pm8$      & $4$   & $8$  \\
  \hline              
  EFF3                   & $22\pm3$   & $10.5\pm0.1$   & $2.1\pm0.2$   & $51\pm11$     & $5$   & $8$ \\
  \hline
  EFF4                   & $31\pm4$   & $11.0\pm0.2$   & $2.8\pm0.4$  & $100\pm22$      & $6$   & $8$ \\
  \hline
  EFF5                   & $152\pm11$ & $11.61\pm0.07$ & $1.95\pm0.16$ & $318\pm37$    & $20$  & $12$ \\
\bottomrule
\end{tabular}
\tablefoot{
\tablefoottext{a}{Full width at half maximum of the line, $\Delta\varv=2\sqrt{2\ln(2)}\sigma_{\varv}$ with $\sigma_{\varv}$ the standard deviation of the Gaussian profile.}\\
\tablefoottext{b}{Using \citet{Lenz1992} for the peak of the line.}
}
\end{table*}

We detected the C$102\alpha$ and C$109\alpha$ CRRLs in four out of the five positions observed.
The averaged CRRL spectra are shown in Figure~\ref{fig:isfspectra}.

To quantify the line properties we decompose the observed spectra using Gaussian line profiles. For the averaged CRRL the broadening effects that produce Lorentzian line wings are negligible, hence the line profiles should be Gaussian \citep[e.g.,][]{Hoang-Bihn1974,Salgado2017b}. During the fitting process we do not restrict the number of velocity components, and we determine how many to include in the fit using the Akaike information criterion \citep[AIC,][]{Akaike1974}. The AIC will have its lowest score when the fit residuals are at a minimum for the smallest number of velocity components necessary to reproduce the spectra. We compute the AIC under the assumption of normally distibuted errors and include a second order bias correction to account for the small number of channels \citep[e.g.,][]{Burnham2004}. For each spectra we assign up to three velocity components, with starting parameters determined by visual inspection. Then, for each spectra we start by fitting a single velocity component, for the brightest feature in the spectrum, and compute the AIC. We increase the number of velocity components in the fit by one until three are included simultaneously. Finally, we select the number of velocity components that gives the lowest AIC. If the difference between the fit with the lowest AIC and the next one is less than $2$, we select the fit with the lowest number of velocity components.

The best fit parameters of the Gaussian fits to the averaged CRRL spectra are presented in Table~\ref{tab:c106a}. All the detected lines are well fit by a single velocity component. The velocity centroid of the lines varies between $10$ and $12$~km~s$^{-1}$ and their full width at half maximum between $\Delta\varv{\approx}1.4$ and $2.8$~km~s$^{-1}$, consistent with observations of the ISF in other tracers of neutral gas. The line widths of the CRRLs are similar to the ones found by \citet{Cuadrado2019} toward the Orion Bar ($\Delta\varv{\approx}2.6\pm0.4$~km~s$^{-1}$), where the CRRLs trace the C$^{+}$/C interface of the PDR, and are not associated with the warm ionized gas.

In EFF1, EFF2 and EFF4 we also detect the HRRL, associated with warm ionized gas, at a velocity of ${\approx}-4$~km~s$^{-1}$ and a line width of $\Delta\varv{\approx}28$~km~s$^{-1}$ (Figure~\ref{fig:isfspectra} left). This line width is significantly broader than that of the CRRLs, and suggests that the gas traced by the HRRLs is warmer than that traced by the CRRLs. The differences between the broad HRRLs and the narrow CRRLs confirms that these lines trace different volumes of gas \citep[see also, e.g.,][]{Goicoechea2015}.

The RRLs from helium (HeRRLs) have a rest frequency which corresponds to a ${\approx}27.4$~km~s$^{-1}$ velocity shift with respect to CRRLs. The HeRRLs, like the HRRLs, are expected to trace warm ionized gas, hence we expect them to have a velocity of ${\approx}-4$~km~s$^{-1}$. This means that the HeRRLs would appear at a velocity of ${\approx}23$~km~s$^{-1}$ in the CRRL spectra. The CRRL spectra in Figure~\ref{fig:isfspectra} show no significant emission at this velocity.

\subsection{Comparison with [CII]}

The averaged CRRLs, [CII] and [$^{13}$CII] $F{=}2\mbox{--}1$ ([$^{13}$CII] hereafter) lines are compared in the central panel of Figure~\ref{fig:isfspectra}. To quantitatively compare [CII] and [$^{13}$CII] with the averaged CRRLs we decompose their line profiles using Gaussian components and the same criteria used for the averaged CRRLs, but including up to six velocity components. The models with the lowest AIC for the [CII] line have three or more velocity components toward each region, while for the [$^{13}$CII] line a single velocity component produces the minimum AIC. The best fit parameters of the Gaussian profile are presented in Table~\ref{tab:13cii} for [$^{13}$CII] and Table~\ref{tab:cii} for [CII].

\begin{figure}[h]
  \resizebox{\hsize}{!}
  {\includegraphics[width=\textwidth]{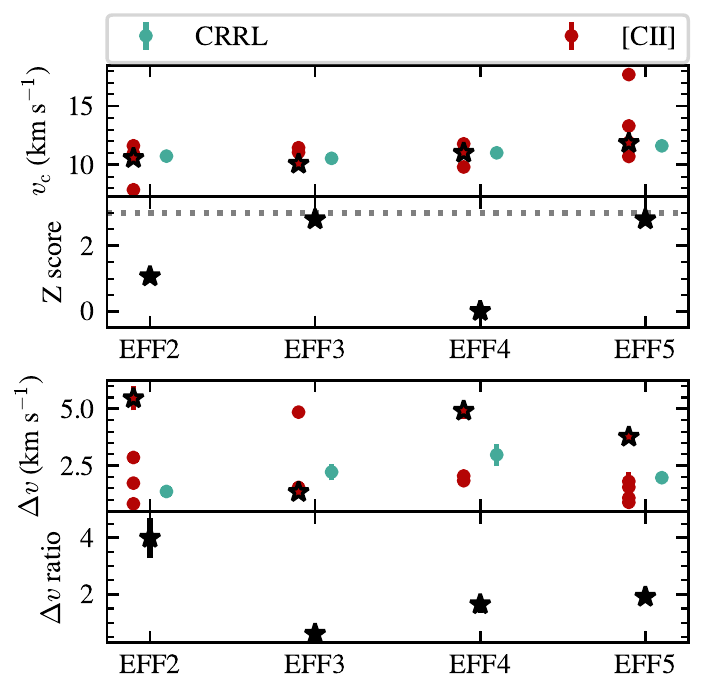}} %
  \caption{Comparison between the velocity centroids, $\varv_{\rm{c}}$, and line widths, $\Delta\varv$, of the [CII] and CRRL line profiles.
  $\varv_{\rm{c}}$ and $\Delta\varv$ are determined from a multicomponent Gaussian fit to the observed spectra (Figure~\ref{fig:isfspectra}).
  The top panel shows the velocity centroids, the upper middle panel shows the difference between the closest velocity centroids in units of its error, Z score, the lower middle panel shows the standard deviation of the Gaussian line profile, and the lower panel shows the ratio $\Delta\varv(\mbox{[CII]})/\Delta\varv(\mbox{[CRRL]})$.
  The starred symbols show the [CII] components that are closest in velocity to the CRRLs.}
  \label{fig:crrlciiprops}
\end{figure}

The averaged CRRLs and [$^{13}$CII] lines have consistent widths and centroids (Figure~\ref{fig:isfspectra}, and Tables~\ref{tab:c106a} and \ref{tab:13cii}), which suggests that both lines trace similar volumes of gas. On the other hand, the composite [CII] line profile is broader and has a different velocity centroid than the averaged CRRLs and the [$^{13}$CII] line. This is particularly evident toward EFF5, where the peak of the averaged CRRLs and the [$^{13}$CII] line is at $11.6\pm0.06$~km~s$^{-1}$, while that of the composite [CII] line is at $10.9\pm0.03$~km~s$^{-1}$. However, when we consider the decomposition of the [CII] profile, there are [CII] components whose centroid agrees with that of the CRRLs and [$^{13}$CII]. A comparison between the velocity centroids and widths of the individual velocity components in the [CII] line and those of the averaged CRRLs is presented in Figure~\ref{fig:crrlciiprops}. This comparison shows that toward each position, there is at least one [CII] velocity component whose velocity centroid agrees with that of the averaged CRRLs (Z score$<3$, upper middle panel in Figure~\ref{fig:crrlciiprops}), but that even for this component the [CII] line is broader than the CRRLs (bottom panel in Figure~\ref{fig:crrlciiprops}).

The difference between the line widths of [CII] and those of the averaged CRRLs and [$^{13}$CII] can be explained either by the [CII] line being optically thick, it tracing multiple velocity components along the line of sight, or it also tracing photoevaporating gas from the surface of the ISF. An optically thick line would appear broader due to the effects of opacity broadening, similar to the effect caused by a blend of multiple velocity components. Photoevaporation would result in a broader line profile as the photoevaporating gas is red and/or blue-shifted with respect to the bulk of the gas. Typical velocity shifts of photoevaporating gas are $0.5\mbox{--}1$~km~s$^{-1}$ \citep{Tielens2005}, in line with the observed velocity shift for EFF5.
It is also possible that a combination of the above scenarios is at play. We return to the width of the [CII] line when we analyze the gas properties (Sect.~\ref{ssec:cgasprops}).

\begin{table*}%[ht]
\caption{\label{tab:13cii} Best fit [$^{13}$CII] $F{=}2\mbox{--}1$ line properties.}%\tablefootmark{$\dagger$}}
\centering
\begin{tabular}{lcccccccccc}
\toprule
Region                  &  $T_{a}$     &  $\varv_{\mathrm{c}}$ &  $\Delta\varv$\tablefootmark{a}  & $\int T_{\mathrm{a}}^{*}dv$ & rms     & S/N\tablefootmark{b} \\
                        & (K)          & (km s$^{-1}$)         & (km s$^{-1}$)                    & (K km s$^{-1}$)             & (K)     &     \\
\midrule
  EFF1                  & $<1$         &                       &                                  &                             &  $0.35$ &      \\
  EFF2                  &  $0.6\pm0.2$ &  $11.2\pm0.1$         & $0.9\pm0.3$                      & $0.6\pm0.3$                 &  $0.33$ &  $3$   \\
  EFF3\tablefootmark{c} &  $0.4\pm0.3$ &  $10.3\pm0.2$         & $0.7\pm0.5$                      & $0.3\pm0.3$                 &  $0.35$ &  $1.5$ \\
  EFF4                  &  $0.7\pm0.1$ &  $10.8\pm0.2$         & $2.1\pm0.5$                      & $1.7\pm0.5$                 &  $0.35$ &  $5$   \\
  EFF5                  &  $2.7\pm0.2$ &  $11.67\pm0.05$       & $1.8\pm0.1$                      & $5.6\pm0.5$                 &  $0.37$ &  $16$  \\
\bottomrule
\end{tabular}
\tablefoot{
\tablefoottext{a}{Full width at half maximum of the line, $\Delta\varv=2\sqrt{2\ln(2)}\sigma_{\varv}$ with $\sigma_{\varv}$ the standard deviation of the Gaussian profile.}\\
\tablefoottext{b}{Using \citet{Lenz1992} for the peak of the line.}\\
\tablefoottext{c}{Marginally detected.}
}
\end{table*}

\begin{figure}[h]
  \resizebox{\hsize}{!}
  {\includegraphics{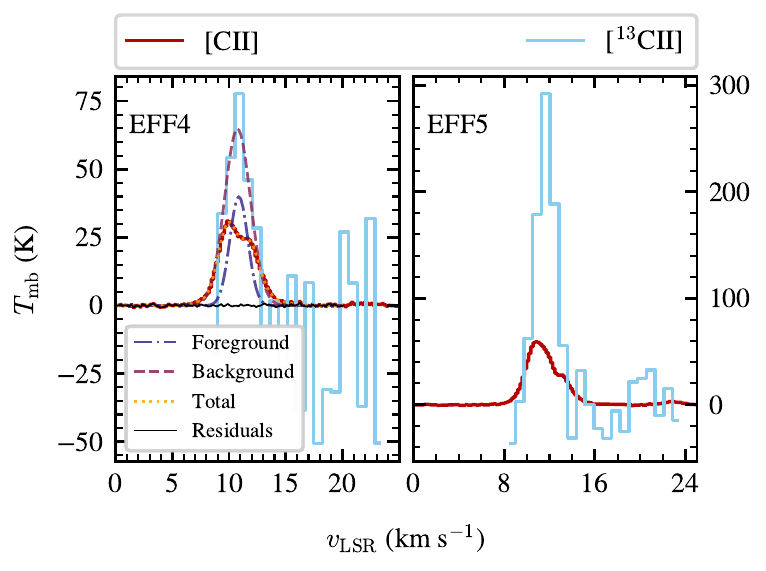}} %
  \caption{Spectra of the [CII] and [$^{13}$CII] $F{=}2\mbox{--}1$ lines toward EFF4 (left) and EFF5 (right).
  The [$^{13}$CII] line has been scaled by the $^{13}$C abundance ([$^{12}$C/$^{13}$C]${=}67$, \citealt{Langer1990}), and the relative strength of its hyperfine component ($0.625$, \citealt{Ossenkopf2013}).
  The scaled [$^{13}$CII] line is brighter than the [CII] line, which suggests that the latter is optically thick, or that there is self-absorption.
  In the left panel we show the results of modeling the [CII] emission considering self-absorption of the background emission by a foreground component.
  In this Figure, we averaged the [$^{13}$CII] line to a channel width of $0.8$~km~s$^{-1}$.
  }
  \label{fig:ciitau}
\end{figure}

In Figure~\ref{fig:ciitau} we show the [CII] line and the [$^{13}$CII] line scaled by the $^{13}$C abundance ([$^{12}$C/$^{13}$C]${=}67$, \citealt{Langer1990}) and the relative strength of its hyperfine component ($0.625$, \citealt{Ossenkopf2013}), toward EFF4 and EFF5. In both cases the scaled [$^{13}$CII] line is brighter than the [CII] line. This indicates that the [CII] line is optically thick, or that there is foreground absorption \citep[see e.g.,][]{Guevara2020}. The prescence of foreground absorption can be inferred from the prescence of absorption features in the IRAC $8~\mu$m image (Figure~\ref{fig:beams}). Cold foreground dust is observed toward EFF3, partially toward EFF2 and EFF4, but not toward EFF1 and EFF5. The prescence of cold $8~\mu$m absorption toward EFF4 suggests that in this region there could be [CII] self-absorption, while the lack of it toward EFF5 suggests that the [CII] line is optically thick.

To determine the effect of [CII] self-absorption toward EFF4 we model the line profile as the superposition of background emission and foreground absorption, both modeled using Gaussian line profiles. We also consider two velocity components to account for the wings of the [CII] line profile. The best fit line profile is presented in Figure~\ref{fig:ciitau}, and its parameters in Table~\ref{tab:cii}. We note that this model has an AIC which is larger than that of the best fit model considering only components in emission ($\Delta$AIC$=6$), which makes it a less likely solution. We use this self-absorbed profile to consider the effects of cold foreground [CII] absorption during our analysis. Toward EFF5 we are not able to find a model with self-absorption that reproduces the observed [CII] line profile ($\Delta$AIC$=140$ with respect to the best fit model without self-absorption).

From the comparison between the [CII] line and the CRRLs, we also note that the [CII] line is detected toward EFF1, the only region not detected in CRRLs. Toward this region the [$^{13}$CII] line is not detected as well. The nondetections of the CRRLs and the [$^{13}$CII] line toward EFF1, also suggests that the C$^{+}$ column density is lower toward this region.

\subsection{Comparison with molecular gas tracers}
\label{ssec:molcomp}

We compare the averaged CRRLs to the HCN$(1\mbox{--}0)$ and C$_{2}$H$(1\mbox{--}0)$ lines in the right panels of Figure~\ref{fig:isfspectra}. There is good agreement between the line profiles, with offsets $<1$~km~s$^{-1}$ in the velocity centroids of the lines. In terms of their widths, the CRRLs are slightly broader than the molecular lines, though their widths are consistent given the uncertainties. The overall agreement between the molecular lines and the CRRLs confirms that the CRRL emission is associated with the neutral gas in the ISF.

Toward EFF1 the HCN$(1\mbox{--}0)$ and C$_{2}$H$(1\mbox{--}0)$ lines are detected at a velocity of ${\approx}12$~km~s$^{-1}$, redshifted with respect to the [CII] line (${\approx}10$~km~s$^{-1}$, Figure~\ref{fig:isfspectra}). The detection of the molecular lines shows that there is molecular gas along the line of sight, and the velocity shift with respect to the warmer neutral material indicates that these gases are not moving at the same speed. The nondetection of the CRRLs along the ${\approx}2$~km~s$^{-1}$ velocity shift between [CII] and HCN suggests that the gas in this region is being dispersed, and that the C$^{+}$ layer is less dense than in other regions.

\begin{figure}[h]
  \resizebox{\hsize}{!}
  {\includegraphics{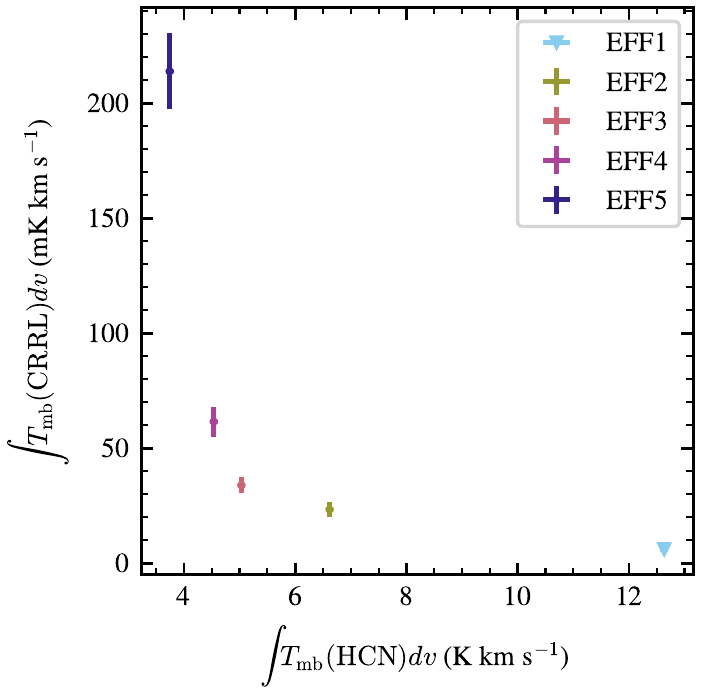}} %
  \caption{Intensity of the averaged CRRLs as a function of the intensity of the HCN$(1\mbox{--}0)$ line.
  The intensities of both lines are anticorrelated. The error bars are $1\sigma$, and the inverted triangle shows the $3\sigma$ upper limit for the CRRL intensity toward EFF1.}
  \label{fig:hcncrrl}
\end{figure}

In Figure~\ref{fig:hcncrrl} we compare the intensity of the averaged CRRLs and the HCN$(1\mbox{--}0)$ line. There, we observe that their intensities are anticorrelated, HCN$(1\mbox{--}0)$ is brighter toward EFF1 and its intensity decreases as we move to EFF5, while the CRRLs are not detected toward EFF1 and are brighter toward EFF5. Naively, we would assume that this anticorrelation can be explained by an increase in the C$^{+}$ column density as we move from EFF1 to EFF5. We return to this anticorrelation in Sect.~\ref{ssec:cgasprops}, once we have derived the properties of the C$^{+}$ layer.

The similarity between the CRRL profile and those of other tracers of the neutral gas in the ISF (Figure~\ref{fig:isfspectra}), confirms that the CRRL emission is associated with the neutral gas in the envelope of the ISF. During the rest of the analysis, we assume that the CRRLs trace the C$^{+}$ layer in the envelope of the ISF.

\section{Analysis}

In this section we derive the electron density, temperature, C$^{+}$ column density and size along the line of sight of the C$^{+}$ layer of the ISF using the averaged CRRLs, [CII] and [$^{13}$CII] $F{=}2\mbox{--}1$ lines. Then, we constrain the ionization fraction deeper into the ISF using the HCN$(1\mbox{--}0)$ and {C$_{2}$H$(1\mbox{--}0)~J{=}1/2\mbox{--}1/2$} lines.

\subsection{C$^{+}$ layer properties}
\label{ssec:c+props}

In this subsection we focus on the properties of the C$^{+}$ layer. We start by describing the model used to derive the properties of this layer, then we present the gas properties derived by comparing the observations to the model predictions, and we end by discussing the uncertainties of this modeling.

\subsubsection{Model}

We model the properties of the C$^{+}$ layer of the PDR on the surface of a molecular cloud as a plane parallel homogenoeus slab of gas.
We assume that throughout the slab carbon is singly ionized, and that it is the dominant donor of free electrons so $x(\mbox{C}^{+})=x(\mbox{e}^{-})=1.5\times10^{-4}$ \citep{Sofia2004}.
This is a good approximation in the C$^{+}$ layer of a PDR \citep[e.g.,][]{Salas2019}.
We also assume that the amount of molecular and atomic gas in the slab is the same, that is, $0.5n_{\mathrm{HI}}+0.5n_{\mathrm{H}_{2}}=n_{\mathrm{e}^{-}}/x(\mbox{e}^{-})$.
Since we assume that the gas properties are homogeneous in the C$^{+}$ layer, we have that the CRRL, [CII] and [$^{13}$CII] $F{=}2\mbox{--}1$ emission originates from the same volume of gas and that their emission fills the telescope beam.

The brightness temperature of the FIR C$^{+}$ line is given by \citep[e.g.,][]{Goldsmith2012}
\begin{equation}
 T_{\mathrm{B}}(\mbox{[CII]})=T^{*}\left(\frac{1}{e^{T^{*}/T^{\mathrm{ex}}}-1}-\frac{1}{e^{T^{*}/T^{\mathrm{bg}}}-1}\right)(1-e^{-\tau}),
\end{equation}
with
\begin{equation}
 \tau=\frac{c^{3}A_{ul}}{8\pi\nu_{\mathrm{CII}}^{3}}\frac{g_{u}}{g_{l}}N(\mbox{C}^{+})\frac{1-e^{-T^{*}/T^{\mathrm{ex}}}}{1+\frac{g_{u}}{g_{l}}e^{-T^{*}/T^{\mathrm{ex}}}}\phi_{\mathrm{FIR}}.
\end{equation}
Here $A_{ul}=2.36\times10^{-6}$~s$^{-1}$ is the spontaneous decay rate for the $158~\mu$m-[CII] line \citep{Wiese2007}, $\nu_{\mathrm{CII}}$ is the frequency of the line ($1900.5369$~GHz for [CII] and $1900.4661$~GHz for [$^{13}$CII] $F{=}2\mbox{--}1$, \citealt{Cooksy1986}), $g_{u}$ and $g_{l}$ are the statistical weights of the upper ($J=3/2$) and lower ($J=1/2$) levels, $T^{*}=\Delta E/k_{\mathrm{B}}=91.25$~K is the energy difference between the upper and lower levels, $T^{\mathrm{ex}}$ is the excitation temperature, $T^{\mathrm{bg}}$ is the temperature of the background continuum and $\phi_{\mathrm{FIR}}$ the line profile in units of km s$^{-1}$.
$T^{\mathrm{ex}}$ is determined from the solution to the detailed balance problem, considering collisions with electrons, atomic and molecular hydrogen, and the rates given in \citet{Goldsmith2012}.
We assume the same excitation temperature for the [CII] and [$^{13}$CII] lines.
To predict the brightness temperature of the [$^{13}$CII] $F{=}2\mbox{--}1$ line we adopt a [$^{12}$C/$^{13}$C] abundance ratio of $67$ \citep{Langer1990}, and a relative strength of $0.625$ \citep{Ossenkopf2013}.

For a transition from upper level $n^{\prime}$ to $n$ the brightness temperature of a CRRL is given by \citep[e.g.,][]{Dupree1974}
\begin{equation}
 T_{\mathrm{B}}(\mbox{CRRL})=\tau^{*}_{\nu}(b_{n^{\prime}}T-b_{n}\beta_{nn^{\prime}}T_{\mathrm{cont}}),
 \label{eq:crrl_ta}
\end{equation}
with
\begin{equation}
 \tau^{*}_{\nu}=\frac{h^{3}\mathrm{e}^{2}\pi}{(2\pi m_{\mathrm{e}^{-}}k_{\mathrm{B}})^{3/2}m_{\mathrm{e}^{-}}c}n^{2}f_{n,n^{\prime}}n_{\mathrm{e}^{-}}N(\mbox{C}^{+})T^{-1.5}e^{\chi_{n}}(1-e^{h\nu/k_{\mathrm{B}}T})\phi_{\mathrm{CRRL}},
 \label{eq:crrl_tau}
\end{equation}
where $b_{n}$ and $\beta_{nn^{\prime}}$ are the departure coefficients, $T_{\mathrm{cont}}$ the brightness temperature of the background continuum, $h$ Planck's constant, $\mathrm{e}$ the electron charge, $k_{\mathrm{B}}$ the Boltzmann constant, $m_{\mathrm{e}^{-}}$ the mass of the electron, $c$ the speed of light, $\chi_{n}=157800n^{-2}T^{-1}$, $f_{n,n^{\prime}}$ the oscillator strength of the transition, $T$ is the gas temperature, $n_{\mathrm{e}^{-}}$ is the density of electrons, $N(\mbox{C}^{+})=n_{\mathrm{C}^{+}}l$ is the column density of ionized carbon, $\phi_{\mathrm{CRRL}}$ the line profile in units of Hz$^{-1}$ and $l$ is the size of the C$^{+}$ layer along the line of sight. To compute the departure coefficients we follow \citet{Salgado2017a}, and compute them for $n_{\mathrm{e}^{-}}$ between $0.01$--$200$~cm$^{-3}$ and $T$ between $10$--$10^{3}$~K. For the oscillator strength we use the approximation by \citet{Menzel1968}, valid for transitions of hydrogenic atoms at large $n$ and small $\Delta n=n^{\prime}-n$.

To determine the gas properties, we compare the averaged CRRLs, [CII] and [$^{13}$CII] $F{=}2\mbox{--}1$ lines and their ratios to the model predictions. We require that the model matches the observations to within $3\sigma$, and the FIR data are convolved to $126\arcsec$ to match the resolution of the radio lines. For the [CII] line we use the brightest velocity component that is consistent with the velocity centroid of the averaged CRRLs and [$^{13}$CII] $F{=}2\mbox{--}1$ lines.

In summary, our model takes as inputs the gas temperature, electron density and C$^{+}$ column density, and from these inputs it predicts the intensities of the $n{=}106$ CRRL, and the FIR lines of C$^{+}$. These intensities are then compared to the observations (Tables~\ref{tab:c106a}, \ref{tab:13cii} and \ref{tab:cii}) to determine the temperature, $n_{\mathrm{e}^{-}}$ and $N(\mbox{C}^{+})$.
From these values we also derive the size of the C$^{+}$ layer along the line of sight, the thermal pressure, and the thermal and nonthermal line widths.

\subsubsection{Gas properties}
\label{ssec:cgasprops}

\begin{table*}%[ht]
\caption{\label{tab:gaspropseff} Gas and line properties derived from the analysis of the CRRLs and the FIR C$^{+}$ lines.}
\centering
\begin{tabular}{lcccccccc}
\toprule
Region & $T^{\mathrm{ex}}$ & $\tau(\mbox{[CII]})$    & $T$          & $n_{\mathrm{e}^{-}}$ & $N(\mbox{C}^{+})$           & $l$   & $\sigma_{\varv,\rm{th}}$  &  $\sigma_{\varv,\rm{nth}}$     \\
       & (K)               &                         & (K)          & (cm$^{-3}$)          & ($\times10^{18}$~cm$^{-2}$) & (pc)  & (km s$^{-1}$)             &  (km s$^{-1}$)                 \\
\midrule  
EFF1   & --                & --            & --           & --               & --            & --            & -- & \\
EFF2   & $49\pm4$          & $4\pm1$       & $50$--$450$   & $\geq0.3$       & $0.05$--$1.6$ & $\leq1.7$     & $0.18$--$0.55$ & $0.16$--$0.55$ \\
EFF3   & --                & --            & --           & --               & --            & --            & -- & \\
EFF4   & $33\pm3$          & $12.8\pm0.8$  & $45\pm5$     & $0.65\pm0.12$    & $2.4\pm0.2$   & $1.1\pm0.2$   & $0.17\pm0.01$  & $1.2\pm0.2$   \\
EFF5   & $63\pm6$          & $10\pm1$      & $55\pm2$     & $0.95\pm0.02$    & $4.5\pm0.2$   & $1.5\pm0.2$   & $0.195\pm0.003$& $0.81\pm0.08$ \\
\bottomrule
\end{tabular}
\end{table*}

The gas properties derived using our model are presented in Table~\ref{tab:gaspropseff}. For EFF4 and EFF5 the gas properties are well constrained, while for EFF2 the lower S/N of the spectra results in larger uncertainties. For regions without a [$^{13}$CII] detection (EFF1 and EFF3), we do not derive gas properties. For EFF4 and EFF5, $n_{\mathrm{e}^{-}}{\sim}0.8$~cm$^{-3}$ which corresponds to a gas density $n_{\rm{H}}{\sim}5\times10^{3}$~cm$^{-3}$.

The thermal pressure for the gas is $p_{\rm{th}}{=}(4.0\pm0.5)\times10^{5}$~K~cm$^{-3}$ and $(2.3\pm0.4)\times10^{5}$~K~cm$^{-3}$ for EFF5 and EFF4, respectively. These thermal pressures are smaller than those in the PDRs of NGC 1977 \citep[$(4\mbox{--}9)\times10^{5}$~K~cm$^{-3}$,][]{Pabst2020} and the Orion Bar \citep[$(2\mbox{--}4)\times10^{8}$~K~cm$^{-3}$,][]{Cuadrado2019}, both of which are exposed to a higher radiation field than the regions studied here.

The C$^{+}$ column densities (Table~\ref{tab:gaspropseff}) are converted to hydrogen column densities using $N(\mbox{H})=N(\mbox{C}^{+})/x(\mbox{C}^{+})$ and are presented in Table~\ref{tab:column}.
The derived $N(\mbox{H})$ is $20\%$ and $80\%$ of the dust derived column density for EFF4 and EFF5, respectively. The warmer region (EFF5) shows a higher fraction of the gas in the C$^{+}$ layer.
This might reflect the fact that EFF5 is exposed to a higher radiation field, hence has a larger C$^{+}$ layer, or that for warmer regions the assumption of isothermal dust along the line of sight biases the dust derived column densities toward lower values.

\begin{table}%[ht]
\caption{\label{tab:column} Column densities.}
\centering
\begin{tabular}{lcccc}
\toprule
Region & $N(\mbox{C}^{+})/x(\mbox{C}^{+})$\tablefootmark{a} & $N(\mbox{H}_{2})$\tablefootmark{b} & $N(\mbox{H})$\tablefootmark{c} & $N(\mbox{H})$\tablefootmark{d} \\
       & \multicolumn{3}{c}{($\times10^{22}$~cm$^{-2}$)}                                                                                                                   \\
\midrule
EFF1   & --          & $4.3$ & --         & $18\pm5$    \\
EFF2   & $0.03$--$1$ & $3.9$ & $4$--$4.9$ & $7.2\pm2$   \\
EFF3   & --          & $2.6$ & --         & $10\pm2$    \\
EFF4   & $1.3\pm0.1$ & $3.1$ & $4.4$      & $6.5\pm1$   \\
EFF5   & $2.7\pm0.1$ & --    & --         & $3.3\pm0.8$ \\
\bottomrule
\end{tabular}
\tablefoot{
\tablefoottext{a}{Converted from $N(\mbox{C}^{+})$ adopting {$x(\mbox{C}^{+})=1.5\times10^{-4}$} \citep{Sofia2004}.}\\
\tablefoottext{b}{From $^{12}\mbox{CO}$ and $^{13}\mbox{CO}$ \citep{Berne2014}.}\\
\tablefoottext{c}{Sum of columns 2 and 3.}\\
\tablefoottext{d}{From the dust SED modeling by \citet{Lombardi2014}.}\\
}
\end{table}

Our model and the observations are consistent with optically thick [CII] lines (Table~\ref{tab:gaspropseff}). For a [CII] optical depth $\tau(\mbox{[CII]}){\approx}10$ the lines would appear to be a factor of two broader due to opacity broadening. This is consistent with the measured line widths (Figure~\ref{fig:crrlciiprops} bottom).

In Figure~\ref{fig:hcncrrl} we observed an anticorrelation between the intensities of the CRRLs and the HCN$(1\mbox{--}0)$ line. The derived C$^{+}$ column densities (Table~\ref{tab:gaspropseff}) increase from EFF2 to EFF5, which would explain the observed relation. As the C$^{+}$ column density increases, so does the CRRL intensity, while the HCN column density and the HCN$(1\mbox{--}0)$ intensity decrease. This implies that, in general, the detection of molecular gas and/or bright [CII] does not guarantee the detection of CRRLs. Given the present observations we detect the C$102\alpha$ and C$109\alpha$ CRRLs from regions with $N(\mbox{C}^{+})\gtrsim5\times10^{16}$~cm$^{-2}$. This means that C-band ($4$--$8$~GHz) observations of CRRLs, where a total of $24$ C$\alpha$ RRLs can be observed, is limited to the study of regions with $N(\mbox{C}^{+})\gtrsim1\times10^{16}$~cm$^{-2}$ using current instruments. This is a loose lower limit, as colder and/or denser regions could be observed at lower column densities.

\subsubsection{Uncertainties}

The largest uncertainties in the derived gas properties arise from (i) the decomposition of the [CII] line profiles into Gaussian components, (ii) the prescence of foreground [CII] absorption and (iii) the assumption that [CII], [$^{13}$CII] and the CRRLs trace the same volume of gas. Here we explore how these effects affect the derived gas properties.

To test how (i), the decomposition of the [CII] line profiles into Gaussian components, affects our results we vary the [CII] intensity. We perform this analysis toward EFF4 and EFF5 as these are the regions with the highest S/N detections. For EFF5 we find that decreasing the intensity of the [CII] line by a factor of $0.8$ increases the electron density to $1.4\pm0.2$~cm$^{-3}$ and lowers the temperature to $50\pm5$~K, while the C$^{+}$ column density remains almost the same, $(4\pm0.3)\times10^{18}$~cm$^{-2}$. And, if we increase the [CII] intensity by a factor of $1.5$ we have $T{=}60\pm7$~K, $n_{\mathrm{e}^{-}}=2.4\pm0.5$~cm$^{-3}$ and $N(\mbox{C}^{+})=(3.8\pm0.4)\times10^{18}$~cm$^{-2}$. For EFF4 if we increase the intensity of the [CII] line by $1.5$ the derived temperature is $65\pm11$~K, $n_{\mathrm{e}^{-}}=5.2\pm1.6$~cm$^{-3}$ and $N(\mbox{C}^{+})=(1.6\pm0.5)\times10^{18}$~cm$^{-2}$. This shows that altough the derived gas properties are sensitive to the decomposition of the [CII] line profiles, the results remain consistent.

The effects of (ii), the presence of foreground [CII] absorption, would depress the background [CII] line profile. Toward EFF4 we could use the results of the [CII] line decomposition (Table~\ref{tab:cii}) when considering self-absorption. However, when we use this decomposition our model only provides a lower limit to the gas temperature and electron density because of the large uncertainties in the best fit line parameters. We place upper limits to the temperature of the gas in the C$^{+}$ layer using the results from the PDR toolbox \citep{Kaufman2006,Pound2008}.
The PDR toolbox predicts the gas temperature at the surface of the PDR given $G_{0}$ and the gas density. For $G_{0}$, we use the values derived from the dust temperature, and for the density the lower limits to the electron density ($n_{\mathrm{H}}\geq n_{\mathrm{e}^{-}}/1.5\times10^{-4}$). Toward EFF4 we have that the gas is exposed to a radiation field of $G_{0}{\approx}35$ and has a density ${>}10^{3}$~cm$^{-3}$, thus the surface temperature is expected to be ${\leq}100$~K. This sets an upper limit to the electron density of $n_{\mathrm{e}^{-}}\leq10$~cm$^{-3}$. Then, from the upper and lower limits, the electron density is constrained to $0.3$~cm$^{-3}\leq n_{\mathrm{e}^{-}}\leq10$~cm$^{-3}$ and the gas temperature to $30$~K$~\leq T\leq100$~K. The lower limits are consistent with the electron density and temperature derived from the emission only analysis, but the upper limits on the density is a factor of $15$ larger. We note that the upper limit is unlikely, since it would imply a gas thermal pressure of $p_{\mathrm{th}}\approx7\times10^{6}$~K~cm$^{-3}$, which is larger than the thermal presure in the PDR of NGC 1977 \citep{Pabst2020}, and would conflict with the nondetection of the C$76\alpha$ line toward EFF4 by \citet{Kutner1985}. When considering self-absorption, the [CII] optical depth is $0.5$ ($1.6$) and $T^{\mathrm{ex}}\approx200$~K ($80$~K) for the background (foreground) components, resulting in a similar column density $N(\rm{C}^{+})\sim10^{18}$~cm$^{-2}$. This analysis shows that when we consider the prescence of [CII] self-absorption, the gas temperature and density could be consistent or larger than the values presented in Table~\ref{tab:gaspropseff}.

To test (iii), assuming that [CII], [$^{13}$CII] and the CRRLs trace the same volume of gas, we compare the predictions of our model to independent observations. We compare to results of the C$76\alpha$ observations that \citet{Kutner1985} obtained with the NRAO $43$~m telescope at $120\arcsec$ resolution. EFF4 and EFF5 show significant overlap with some of their observations (Figure~\ref{fig:beams} right). Close to EFF5 they report a line peak of $T_{\rm{R}}^{*}=29\pm10$~mK, a line width of $\Delta\varv=0.8$~km~s$^{-1}$ and an intensity $T_{\rm{R}}^{*}d\varv=54\pm9$~mK for the line at $\varv_{\mathrm{c}}=11.5$~km~s$^{-1}$. Given the values and uncertainties they report, we find that the line intensity and its uncertainty are underestimated. Here we compute an intensity of $T_{\rm{R}}^{*}d\varv=58\pm20$~mK, given the line peak and width (adopting an error on the line width of $0$~km~s$^{-1}$). Toward EFF5 we predict a C$76\alpha$ intensity of $95\pm2$~mK~km~s$^{-1}$, consistent with the value we derived. Close to EFF4 they report a nondetection of the C$76\alpha$ line with a $3\sigma$ upper limit of $T_{\rm{R}}^{*}\leq60$~mK, for which we predict $T_{\rm{R}}^{*}=23\pm2$~mK. The agreement between our model predictions and the C$76\alpha$ observations suggests that modeling the FIR [CII] and [$^{13}$CII] lines as arising from the same volume as the CRRLs is a reasonable assumption.

An additional test to our model comes from comparing the derived gas properties to those derived by \citet{Kutner1985}. \citet{Kutner1985} derive a temperature of $T{\approx}35$~K and $n_{\mathrm{e}^{-}}{\approx}3$~cm$^{-3}$ from the C$110\alpha$ and C$76\alpha$ line ratio at $396\arcsec$ resolution on the edge of the HII region NGC 1977, encompassing EFF5. They estimate a factor of two uncertainty in the gas properties, which could explain the difference in temperature, but not the factor of three difference in $n_{\mathrm{e}^{-}}$. The remaining difference in density can be explained by the different departure coefficients used ($b_{n}\approx0.05$ using \cite{Walmsley1982} and $b_{n}\approx0.07$ using \cite{Salgado2017a} for $n_{\mathrm{e}}=10$~cm$^{-3}$ and $T=20$~K) and the use of an approximation to compute the C$n\alpha$ optical depth by \citet{Kutner1985} (they approximate the last term in parenthesis in Eq.~\ref{eq:crrl_tau}, which is not a good approximation for $n<100$). These differences would lead us to derive densities that are a factor of two lower than those of \citet{Kutner1985}, bringing our results into agreement.

For a fixed temperature and size of the region along the line of sight the electron density is proportional to the square root of the brightness of the CRRLs. Thus, if the observed lines are beam diluted at 126\arcsec\ resolution then the derived electron densities are underestimated. Conversely, for a fixed line brightness $n_{\mathrm{e}^{-}}$ increases with temperature and decreases for larger column densities. If the CRRL and [CII] emission have a similar spatial extent (as suggested by the observations of \citealt{Wyrowski1997} on $0.02$~pc scales toward the Orion Bar), then beam dilution is not significant at 126\arcsec.

Our observations sample the gas at 126\arcsec\ resolution ($0.25$~pc at the distance of Orion~A). At this resolution we expect that multiple clumps and structures will be unresolved, hence the derived gas properties are an average of the properties of individual clumps and filaments within the beams. Additionally, since the line intensities of the CRRLs and the [CII] line have different dependencies on the gas temperature and density, they will trace different portions of the C$^{+}$ layers within the telescope's beam. If the temperature and density gradients are large in the C$^{+}$ layer, then using an homogeneous model will result in biased gas properties. In particular, the gas temperatures and densities could be biased toward higher values. The comparison between the predictions of our model and the results of \citet{Kutner1985} suggests that in the regions studied these gradients are small enough that this effect does not significantly affect our results.
Using a model that takes into account the temperature and density gradients can reduce these biases \citep[e.g.,][]{Salas2019}. Another consequence of the differences in how the radio and FIR C$^{+}$ lines respond to temperature and density, is that on scales $<0.02$~pc we expect to start observing differences between their spatial distributions. If the differences between the spatial distributions of the CRRLs and the [CII] line are resolved, then the use of a homogeneous model for the C$^{+}$ layer is also expected to produce biased results.

Lastly, we consider the influence of the choice of the molecular gas fraction, and the principal quantum number adopted for the averaged CRRLs. We estimate that the molecular gas fraction has a small effect on the derived gas proprieties. If we use $0.9n_{\mathrm{HI}}+0.1n_{\mathrm{H}_{2}}=n_{\mathrm{e}^{-}}/x(\mbox{e}^{-})$, the derived gas density and temperature are lower by $8\%$ and $13\%$, respectively. Increasing the fraction of molecular gas to $90\%$ has no noticeable effect on the derived gas properties. For the principal quantum number, if we used $n=109$ for the averaged CRRLs, then the gas temperature increases by $10\%$ and there is no noticeable change in the electron density.

\subsection{Gas traced by HCN}
\label{ssec:hcngas}

Here we focus on the properties of the gas traced by HCN. First we assess whether the CRRLs and HCN$(1\mbox{--}0)$ trace the same volume of gas, and then we proceed to derive the ionization fraction of the gas using molecular lines.

\subsubsection{HCN \& CRRLs}

The comparison between the HCN$(1\mbox{--}0)$ line and the CRRLs (Sect.~\ref{ssec:molcomp}) shows that both trace the ISF. However, from our data alone it is not clear whether both lines probe the same volumes of gas. Here, we use results from the literature and the gas properties derived for the C$^{+}$ layer to answer this.

\citet{Wyrowski2000} observed NGC 2023 at higher spatial resolution (${\approx}10\arcsec$) in CRRLs and HCN$(1\mbox{--}0)$. Their observations show differences in the spatial distribution of both tracers, with CRRLs tracing gas closer to the exciting star than the HCN$(1\mbox{--}0)$ line. Although the properties of NGC 2023 are different from those of the OMC-2/3, their results suggest that both lines do not trace the same volumes of gas.

Another comparison between CRRLs and HCN$(1\mbox{--}0)$ comes from the gas temperature. \citet{Hacar2020} showed that the HCN/HNC intensity ratio, for the $J{=}1\mbox{--}0$ lines, is an accurate probe of the gas temperature, and that the temperatures derived from this ratio are similar to the ones derived from the lines of ammonia \citep{Friesen2017} and analysis of the FIR emission from dust \citep{Lombardi2014}. In Table~\ref{tab:moltemp} we present the gas temperature derived from the ammonia lines \citep{Friesen2017}, the HCN/HNC ratio \citep{Hacar2020} and that derived for the C$^{+}$ layer (Sect.~\ref{ssec:c+props}). The temperatures derived from the ammonia lines and the HCN/HNC ratio are consistent, while the C$^{+}$ layer is warmer.
This confirms that in the OMC-2/3 the HCN$(1\mbox{--}0)$ line probes a different volume of gas than the CRRLs.

\begin{table}%[ht]
\caption{\label{tab:moltemp} Gas temperature.}
\centering
\begin{tabular}{lccc}
\toprule
Region & $T$(NH$_{3}$)\tablefootmark{a} & $T$(HCN/HNC)\tablefootmark{b} & $T$(C$^{+}$)\tablefootmark{c}\\
       & \multicolumn{3}{c}{(K)}                             \\
\midrule
EFF1   & $23\pm1$     & $23$ & --          \\
EFF2   & $23\pm1$     & $23$ & $50$--$450$ \\
EFF3   & $16.9\pm0.9$ & $18$ & --          \\
EFF4   & $21\pm1$     & $26$ & $45\pm5$    \\
EFF5   & $29\pm2$     & $29$ & $55\pm2$    \\
\bottomrule
\end{tabular}
\tablefoot{
% \tablefoottext{a}{$\times10^{22}$~cm$^{-2}$.}\\
\tablefoottext{a}{Derived from the NH$_{3}$ $(1,1)$, $(2,2)$ and $(3,3)$ lines \citep{Friesen2017}.}\\
\tablefoottext{b}{Derived from the HCN/HNC intensity ratio of the $J{=}1\mbox{--}0$ lines \citep{Hacar2020}.}\\
\tablefoottext{b}{Derived from the [CII], [$^{13}$CII] and CRRLs (Sect.~\ref{ssec:c+props}).}\\
}
\end{table}

\begin{figure}[h]
  \resizebox{\hsize}{!}
  {\includegraphics{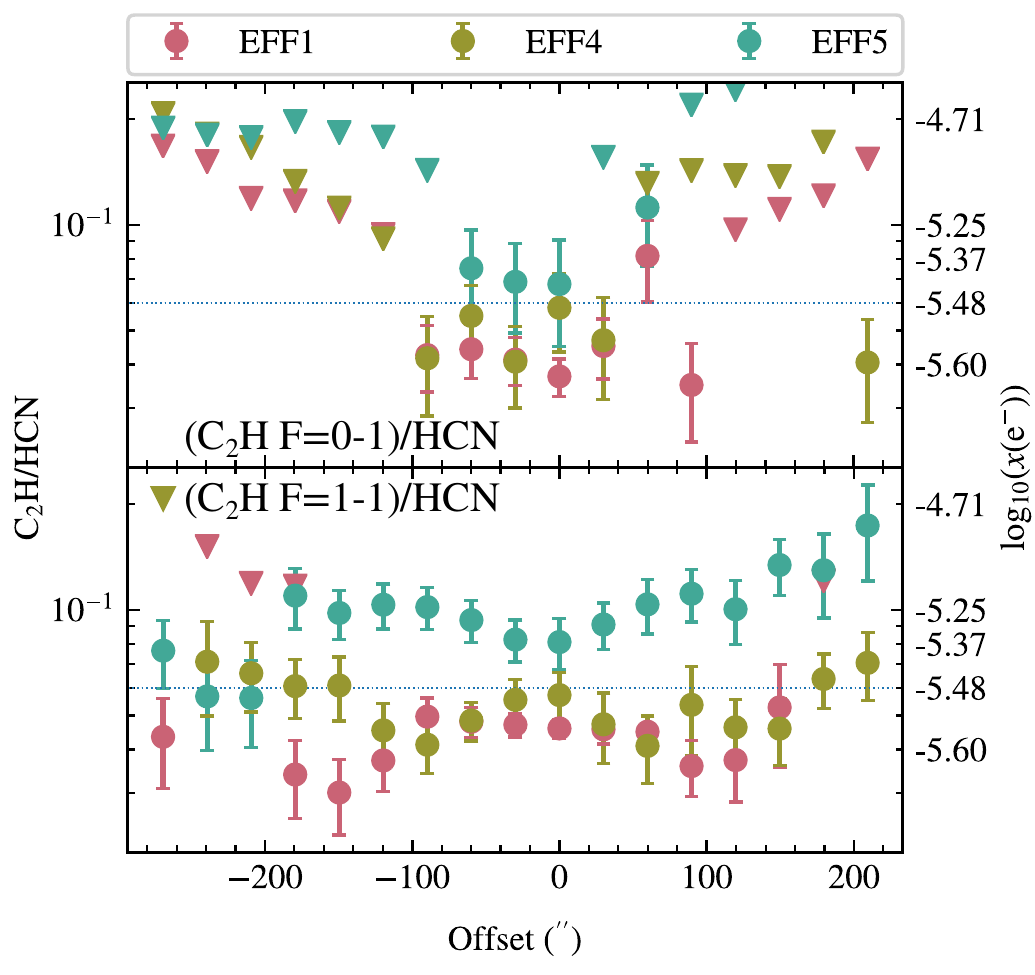}} %
  \caption{Ratio between C$_{2}$H$(1\mbox{--}0)~J{=}1/2\mbox{--}1/2$ and HCN$(1\mbox{--}0)$ for EFF1, EFF4 and EFF5.
           We use this ratio to put an upper limit to $x(\rm{e}^{-})$ in the gas traced by HCN using the models of \citet{Bron2021} (right axis).
           Top: Ratio between C$_{2}$H$(1\mbox{--}0)~J{=}1/2\mbox{--}1/2~F{=}0\mbox{--}1$ and HCN$(1\mbox{--}0)$ for EFF1, EFF4 and EFF5.
           Bottom: Ratio between C$_{2}$H$(1\mbox{--}0)~J{=}1/2\mbox{--}1/2~F{=}1\mbox{--}1$ and HCN$(1\mbox{--}0)$ for EFF1, EFF4 and EFF5.
           The intensity of the $F{=}1\mbox{--}1$ line is scaled to that of the $F{=}0\mbox{--}1$ line by assuming a 2:1 ratio.
           The ratios are computed by averaging over a slice of $2\arcmin$ in the DEC direction centered in the regions.
           The offsets are shown along the RA direction and are referenced to the center of the region (Table~\ref{tab:effobs}).
           Inverted triangles show $3\sigma$ upper limits.}
  \label{fig:xe}
\end{figure}

\subsubsection{Ionization fraction of the molecular gas}

To constrain the ionization fraction of the gas traced by the HCN$(1\mbox{--}0)$ line we use the analytical fits of \citet{Bron2021}. These analytical fits provide $x(\rm{e}^{-})$ as a function of the ratios of the intensity or the column density of different tracers. The analytical fits are derived by varying the physical conditions (e.g., temperature, density, radiation field, $A_{V}$, CRIR) used as input for the chemical network of \citet{Roueff2015}. This produces column densities for the species being considered which are then used to estimate line intensities using the non-LTE radiative transfer program RADEX \citep{vanderTak2007}. We use their results for dense cold clouds, as the visual extinction is ${>}7$ toward the regions studied.

Here we use the intensity ratios of the C$_{2}$H$(1\mbox{--}0)~J{=}1/2\mbox{--}1/2~F{=}0\mbox{--}1$ and HCN$(1\mbox{--}0)$ lines. We use this ratio, as the lines have been observed simultaneously \citep{Melnick2011}, and variations in this intensity ratio can be explained by variations in $x(\rm{e}^{-})$ \citep{Bron2021}. One major caveat of using these lines is that they do not trace the same volume of gas. As shown by observations of the Orion Bar by \citet{vanderWiel2009}, HCN and C$_{2}$H do not trace the same volumes of gas, with C$_{2}$H tracing gas closer to the source of radiation (the Trapezium stars). Since the analytical expressions derived by \citet{Bron2021} are based on a one zone model (i.e., there is no spatial information), this implies that the ionization fractions derived from this ratio will be an average of the ionization fractions in the volumes probed by C$_{2}$H and HCN, and thus can only be used as upper limits to $x(\rm{e}^{-})$ in the gas probed by the HCN$(1\mbox{--}0)$ line.

The line ratios for EFF1, EFF4 and EFF5 are presented in Figure~\ref{fig:xe}. The ratios are consistent with being constant along the RA direction and have values $\approx0.06$, which corresponds to $x(\rm{e}^{-})\lesssim3\times10^{-6}$ in the gas probed by HCN. This upper limit to the ionization fraction is a factor of a few larger than the typical values found toward dense cores $x(\rm{e}^{-})\sim10^{-8}\mbox{--}10^{-6}$ \citep[e.g.,][]{Caselli1998,Bergin1999}.

Away from the spine of the ISF the C$_{2}$H$(1\mbox{--}0)~J{=}1/2\mbox{--}1/2~F{=}0\mbox{--}1$ line is not detected. In these regions we use the C$_{2}$H$(1\mbox{--}0)~J{=}1/2\mbox{--}1/2~F{=}1\mbox{--}1$ line to estimate the ionization fraction by assuming a ratio of $2$ between the $F{=}1\mbox{--}1$ and $F{=}0\mbox{--}1$ lines. The ratio shows small variations along the slices and is consistent with the ionization fraction observed toward the center of the ISF (Figure~\ref{fig:xe} bottom). These results show that the average ionization fraction remains relatively high ($x(\rm{e}^{-}){\approx}3\times10^{-6}$) in the lower column density regions away from the spine of the ISF. If we use the analytical expressions for translucent clouds then the derived ionization fractions would be a factor of $2.5$ larger.

Additional information on the ionization fraction of the molecular gas can be obtained from observations of deuterated molecules. Assuming steady-state chemistry it can be shown that $R_{D}\equiv$[DCO$^{+}$]/[HCO$^{+}$]${\propto}x(\rm{e}^{-})^{-1}$ \citep[e.g.,][]{Caselli1998,Caselli2002}. Around EFF5, \citet{Kutner1985} reports a DCO$^{+}$/H$^{13}$CO$^{+}$ intensity ratio of ${\approx}0.5$ (their Figure~10). For comparison, \citet{Bergin1999} finds DCO$^{+}$/H$^{13}$CO$^{+}$ intensity ratios ${\approx}(1.3\mbox{--}3.8)$ toward massive cores in Orion~A and Orion~B, which they use to derive an average ionization fraction $x(\rm{e}^{-})=(7.7\pm0.1)\times10^{-8}$ for their sample of massive cores. Then, the results of \citet{Kutner1985} suggest that toward EFF5 the ionization fraction is higher than in the massive cores observed by \citet{Bergin1999}. In either case, the properties of the molecular gas in EFF5 are reminiscent of those of a PDR rather than a massive core, and this higher level of ionization may reflect the photo-ionization of trace atoms with low ionization potentials such as S, Mg and Fe by penetrating UV photons.

\section{Discussion}

\subsection{Velocity structure}
\label{ssec:velocity}

High spatial (${\approx}4\arcsec$) and spectral resolution observations of the ISF using optically thin tracers show the presence of multiple velocity components \citep[e.g.,][]{Hacar2018,Zhang2020}. These velocity components reflect a superposition of gas at different velocities along the line of sight, and are associated with the substructures, or fibers, that make up the ISF. At the spatial resolution of our CRRL observations these velocity components are likely to be blended \citep{Hacar2018}, which means that the measured line widths are likely overestimated.

Due to its high collisional excitation rates the FIR [CII] line probes a variety of physical conditions (neutral and ionized gas). The decomposition of the [CII] line profiles shows that $60\%$--$70\%$ of the [CII] intensity is associated with the gas traced by CRRLs in the C$^{+}$/C interface. The remaining velocity components in the [CII] line could be associated with photoevaporating gas from the surface of the molecular cloud. This gas would be warmer and less dense than the components associated with $5$~GHz CRRL emission, resulting in the lack of CRRL detections from these velocity components.

\subsection{Turbulence}

The line widths contain information about the turbulence of the gas. Once the gas temperature is known, the nonthermal contribution to the line widths can be estimated from $\sigma_{\varv}^{2}=\sigma_{\varv,\rm{th}}^{2}+\sigma_{\varv,\rm{nth}}^{2}$, with $\sigma_{\varv,\rm{th}}^{2}=2k_{\rm{B}}T/m$ and $m$ the mass of the atom/molecule.

Toward EFF4 and EFF5, the line widths are dominated by nonthermal broadening (Table~\ref{tab:gaspropseff}). The sound speed, $c_{\mathrm{s}}=\sqrt{k_{\mathrm{B}}T/\mu m_{\mathrm{H}}}$, of the neutral gas ($\mu=1.6$) is $0.48\pm0.03$~km~s$^{-1}$ and $0.53\pm0.01$~km~s$^{-1}$ toward EFF4 and EFF5, respectively. From this we have a sonic Mach number $\mathcal{M}_{s}=\sigma_{\mathrm{nth}}/c_{\mathrm{s}}=2.6\pm0.5$ (EFF4) and $1.5\pm0.1$ (EFF5). We compare this to the sonic Mach number inferred from the ammonia lines \citep{Friesen2017}. Toward EFF4 they report $\sigma_{\varv}=0.44\pm0.02$~km~s$^{-1}$ and $T_{\rm{k}}=21\pm1$~K. This implies that $\sigma_{\varv,\rm{th}}=0.101\pm0.003$~km~s$^{-1}$, $\sigma_{\varv,\rm{nth}}=0.43\pm0.02$~km~s$^{-1}$, and $c_{\mathrm{s}}=0.275\pm0.007$~km~s$^{-1}$ ($\mu=2.33$), so $\mathcal{M}_{s}=1.55\pm0.09$. And, toward EFF5 they find $\sigma_{\varv}=0.35\pm0.03$~km~s$^{-1}$ and $T_{\rm{k}}=29\pm2$~K, for the component at $11.18\pm0.03$~km~s$^{-1}$. For the same component this implies, $\sigma_{\varv,\rm{th}}=0.120\pm0.005$~km~s$^{-1}$, $\sigma_{\varv,\rm{nth}}=0.32\pm0.03$~km~s$^{-1}$, and $c_{\mathrm{s}}=0.32\pm0.01$~km~s$^{-1}$, so $\mathcal{M}_{s}=1.0\pm0.1$. The sonic Mach numbers inferred for the C$^{+}$ layer are in rough agreement with those derived from the ammonia lines. Magnetohydrodynamic (MHD) simulations of the ISM suggest that the sonic Mach number scales with density as $\mathcal{M}_{s}\propto n^{1/2}$ up to densities of $n\approx30$~cm$^{-3}$ and then flattens \citep[e.g.,][]{Kritsuk2004,Kritsuk2017}. Our results are consistent with the scaling derived from MHD simulations, however, our observations likely do not properly resolve the velocity structure of the gas and might result in broader line profiles due to line blending (Sec.~\ref{ssec:velocity}). CRRL observations with a higher spatial resolution could probe how turbulence changes from the envelope of the molecular cloud to its core.

From the nonthermal line widths we estimate the nonthermal pressure, $p_{\rm{nth}}{=}n_{\rm{H}}\mu m_{\mathrm{H}}\sigma_{v,\rm{nth}}^{2}$.
Toward EFF4 and EFF5 the nonthermal pressures are $(1.3\pm0.5)\times10^{6}$~K~cm$^{-3}$ and $(8\pm1)\times10^{5}$~K~cm$^{-3}$, respectively.
The nonthermal pressure is larger than the thermal pressure of the gas, $p_{\rm{th}}{=}(1.9\pm0.4)\times10^{5}$~K~cm$^{-3}$ and $(3.5\pm0.1)\times10^{5}$~K~cm$^{-3}$ for EFF4 and EFF5, respectively.
This suggests that the gas is supported by turbulence.

\subsection{Excitation of HCN}

Here we explore the importance of collisional excitation by electrons, cosmic ray heating and radiative trapping on the intensity of the HCN$(1\mbox{--}0)$ line. To estimate how the HCN$(1\mbox{--}0)$ intensity changes when considering collisions with electrons we use RADEX \citep{vanderTak2007} and the collisional rate coefficients calculated by  \citet{Faure2007} and \citet{Dumouchel2010} for collisions with H$_{2}$ and e$^{-}$, respectively. To estimate the effect of cosmic-ray heating we derive the CRIR using simplified analytical results and then determine the gas equilibrium temperature by balancing the gas heating and cooling rates through the use of the Derive the Energetics and SPectra of Optically Thick Interstellar Clouds code \citep[DESPOTIC,][]{Krumholz2014}. DESPOTIC includes approximations for the dominant gas heating and cooling mechanisms and uses a one-zone model to determine the gas equilibrium temperature. To compute the line cooling it uses the escape probability formalism. We assess the effect of radiative trapping by estimating the HCN$(1\mbox{--}0)$ optical depth from its fine structure components. 

To investigate the HCN intensity using RADEX we adopt a gas temperature of $23$~K, consistent with the temperature of the molecular gas toward EFF4 and EFF5 (Table~\ref{tab:moltemp}), and assume a plane parallel geometry. From the derived path lengths (Table~\ref{tab:gaspropseff}) and the total column densities (Table~\ref{tab:column}) we estimate gas densities of $(1.9\pm0.5)\times10^{4}$~cm$^{-3}$ (EFF4) and $(7\pm2)\times10^{3}$~cm$^{-3}$ (EFF5). Adopting our upper limit to $x(\mathrm{e}^{-})$, these densities correspond to $n_{\mathrm{e}^{-}}=0.048$~cm$^{-3}$ (EFF4) and $0.021$~cm$^{-3}$ (EFF5). For these conditions, we compute the HCN$(1\mbox{--}0)$ intensity with and without electrons as collisional partners. Including collisions with electrons produces a $12\%$ increase in the HCN$(1\mbox{--}0)$ intensity in both cases, almost independent of the adopted HCN column density. The effect is more pronounced at lower temperatures, for example, at $10$~K the increase is $20\%$.

As discussed in Sect.~\ref{ssec:hcngas}, the properties of the molecular gas toward EFF5 seem to be closer to those of the gas in a PDR than the gas in a molecular core. Past the C$^{+}$ layer of a PDR, the ionization fraction of the gas is set by the photo-ionization of trace atoms, and, past the PDR boundary, in the molecular core by cosmic-ray ionization. It is only at an $A_{V}\sim10$, that the ionization of trace atoms (particularly of sulfur and magnesium) can result in $x(\mathrm{e}^{-}){\sim}10^{-6}$, consistent with our upper limits for the HCN traced gas. If the HCN gas is probing regions at an $A_{V}\sim10$ then, we do expect that cosmic-ray heating will be the dominant heating mechanism, as the attenuation of the FUV field this deep in the PDR results in a weak contribution from photoelectric heating.

The equilibrium temperature of the gas is set by the balance between heating and cooling. For the gas in the C$^{+}$ layer the dominant heating mechanism is photoelectric heating, and the cooling is mainly through the [CII] line. Deeper into the cloud ($A_{V}>7$), the heating is mainly due to cosmic ray heating, and cooling is dominated by CO emission. The gas heating and cooling rates will depend on the gas density, the CRIR, and the abundances of the dominant cooling elements. Thus, to estimate the equilibrium temperature, and the effect of cosmic-ray heating, we need to estimate the CRIR.

If cosmic-ray ionization is the dominant ionization mechanism, then the ionization fraction is set by the balance between cosmic-ray ionization and recombination, so $x(\mathrm{e}^{-}){=}C_{i}\zeta_{\mathrm{H}_{2}}^{1/2}n_{\mathrm{H}_{2}}^{-1/2}$. Based on an analytical model, \citet{McKee1989} derive $C_{i}=3.2\times10^{3}$~cm$^{-3/2}$~s$^{1/2}$. For a gas density of $1.6\times10^{4}$~cm$^{-3}$ and an ionization fraction ${\leq}3\times10^{-6}$ this corresponds to $\zeta_{\mathrm{H}_{2}}{\leq}10^{-14}$~s$^{-1}$. We also estimate the CRIR from the $R_{H}\equiv x(\mbox{HCO}^{+})/x(\mbox{CO})$ ratio. \citet{Ohashi2014} observed H$^{13}$CO$^{+}$ toward the millimeter bright source MMS 2 \citep[or TUKH003 in][]{Tatematsu1993}, a dense core that contains a MIR binary system \citep{Nielbock2003} within the EFF4 area, and find $N(\mbox{H}^{13}\mbox{CO}^{+})=4.1\times10^{12}$~cm$^{-2}$ over a $31\farcs5$ beam. Toward the same source $N(^{13}\mbox{CO})=6.7\times10^{16}$~cm$^{-2}$ \citep{Berne2014}, so $R_{H}\equiv$[HCO$^{+}$]/[CO]$=6\times10^{-5}$. Using Eq.~(2) of \citet{Caselli1998} we find $\zeta\leq3\times10^{-15}$~s$^{-1}$ for a temperature of $30$~K. Both upper limits on the CRIR are consistent with the nondetection of HRRLs associated with the cold neutral gas toward EFF4, which implies $\zeta_{p}\leq8\times10^{14}$~s$^{-1}$ \citep{Neufeld2017}. A more precise determination of the CRIR would require a measurement of the ionization fraction for the molecular gas.

We use the upper limits to the CRIR to estimate the gas equilibrium temperature. To model the gas probed by HCN we adopt a density of $1.6\times10^{4}$~cm$^{-3}$, a column density of $6.5\times10^{22}$~cm$^{-2}$, a CRIR of $3\times10^{-15}$~s$^{-1}$ and the parameters for a giant molecular cloud in the Milky Way \citep{Krumholz2014}. We find an equilibrium temperature of ${\sim}45$~K, which is larger than the temperature of the molecular gas inferred from the HCN/HNC ratio and NH$_{3}$ lines (Table~\ref{tab:moltemp}). This suggests that the CRIR is well below our upper limit. If we adopt a standard CRIR of $2.5\times10^{-17}$~s$^{-1}$, then the equilibrium temperature is ${\sim}8$~K. To match the temperature of the molecular gas, $23$~K, the CRIR should be $5\times10^{-16}$~s$^{-1}$, comparable to the values inferred toward dense cores \citep[e.g.,][]{Caselli1998}. An increase in the gas temperature from $8$~K to $23$~K results in a factor of $2.6$ increase in the HCN$(1\mbox{--}0)$ intensity.

Radiative trapping becomes important when the gas is optically thick. \citet[][]{Melnick2011} estimate that the HCN$(1\mbox{--}0)$ optical depth is $\leq1.5$, and less than unity for $90\%$ of the regions where the hyperfine components are detected. An optical depth of $1.5$ results in an escape probability of $0.2$ for a plane parallel slab, which lowers the critical density of the transition by the same amount \citep[e.g.,][]{Shirley2015}. Since the highest optical depths are measured toward the high column density regions, radiative trapping does not explain the extended HCN emission.

These results suggest that toward the spine of the ISF, the HCN$(1\mbox{--}0)$ emission is caused by the presence of warm HCN. This is similar to the findings of \citet{Barnes2020}, who find that the HCN$(1\mbox{--}0)$ intensity per unit column density is higher in the warmer regions around W49. Our analysis suggest that in these regions the gas temperature is set by cosmic-ray heating with a CRIR ${\sim}5\times10^{-16}$~s$^{-1}$.

\section{Summary}

We present observations of the C$102\alpha$ and C$109\alpha$ lines toward five positions along the OMC-2 and OMC-3, with detections in four of these positions. We average these CRRLs to increase the S/N of the spectra during our analysis. Below, we summarize our findings.

\begin{itemize}
 \item The optically thin CRRLs are detected at velocities of $\varv_{\rm{LSR}}{\approx}10$--$12$~km~s$^{-1}$, consistent with the velocity of the ISF in other tracers, such as $158~\mu$m-[CII], CO, HCN, and C$_{2}$H. The width of the line is $\Delta\varv{\approx}2$~km~s$^{-1}$ which confirms that the line traces cold gas ($T<100$~K) in the C$^{+}$/C interface of the ISF.
 \item  We compare the intensities of the CRRLs, [$^{13}$CII] $F{=}2\mbox{--}1$ and [CII] lines to the predictions of a homogeneous model of the C$^{+}$/C interface. For the velocity components detected in the three lines with a SNR ${\geq}5$ we find electron densities of $n_{\mathrm{e}^{-}}{=}0.65\pm0.12$~cm$^{-3}$ (EFF4) and $0.95\pm0.02$~cm$^{-3}$ (EFF5), and gas temperatures of $T{=}45\pm5$~K (EFF4) and $55\pm2$~K (EFF5). These electron densities correspond to a gas density of ${\sim}5\times10^{3}$~cm$^{-3}$, and, combined with the gas temperature, a gas thermal pressure $p_{\mathrm{th}}{\approx}4\times10^{5}$~K~cm$^{-3}$.
 \item For these same regions the temperature of the molecular gas is ${\approx}25$~K, based on the HCN/HNC ratio and the NH$_{3}$ lines. The difference in the temperatures derived from the C$^{+}$ lines and the molecular lines leads us to conclude that they do not trace the same volumes of gas.
 \item  We use the C$_{2}$H$(1\mbox{--}0)~J{=}1/2\mbox{--}1/2$ and HCN$(1\mbox{--}0)$ lines to constrain the ionization fraction in the denser regions of the molecular cloud using the models of \citet{Bron2021}. We find $x(\rm{e}^{-})\leq3\times10^{-6}$.
 \item  We use the ionization fraction of the molecular gas to estimate how collisional excitation with electrons influences the HCN emission. We find that collisions with electrons contribute ${\leq}12\%$ to the line intensity.
\end{itemize}

\noindent
Future observations of CRRLs and HCN in low column density regions will provide further insight into the excitation of HCN in the extended envelopes of molecular clouds.

\begin{acknowledgements}
      The authors thank Gary Melnick and Volker Tolls for sharing their HCN and C$_{2}$H data cubes.
      M.~R.~R. would like to thank B.~Winkel, C.~Kasemann, A.~Kraus and the telescope staff for their support with the setup, conduction and reduction of the observations with the Effelsberg 100m radio telescope, as well as A. Jacob and Y. Gong for useful discussions on the removal of baseline ripples in the spectra observed with Effelsberg.
      This work was based on observations with the 100-m telescope of the MPIfR (Max-Planck-Institut f\"{u}r Radioastronomie) at Effelsberg.
      This research made use of Astropy, a community-developed core Python package for Astronomy \citep{Astropy2013,Astropy2018}, matplotlib, a Python library for publication quality graphics \citep{Hunter2007}, LMFIT, a nonlinear least-square minimization and curve-Fitting package for Python \citep{newville_2014_11813}, and CRRLpy, a Python package for the analysis of RRL observations \citep{crrlpy}.
\end{acknowledgements}

\bibliographystyle{aa}
\bibliography{refs}

\begin{thebibliography}{101}
\expandafter\ifx\csname natexlab\endcsname\relax\def\natexlab#1{#1}\fi

\bibitem[{{Akaike}(1974)}]{Akaike1974}
{Akaike}, H. 1974, IEEE Transactions on Automatic Control, 19, 716

\bibitem[{{Astropy Collaboration} {et~al.}(2018){Astropy Collaboration},
  {Price-Whelan}, {Sip{\H{o}}cz}, {G{\"u}nther}, {Lim}, {Crawford}, {Conseil},
  {Shupe}, {Craig}, {Dencheva}, {Ginsburg}, {VanderPlas}, {Bradley},
  {P{\'e}rez-Su{\'a}rez}, {de Val-Borro}, {Aldcroft}, {Cruz}, {Robitaille},
  {Tollerud}, {Ardelean}, {Babej}, {Bach}, {Bachetti}, {Bakanov}, {Bamford},
  {Barentsen}, {Barmby}, {Baumbach}, {Berry}, {Biscani}, {Boquien}, {Bostroem},
  {Bouma}, {Brammer}, {Bray}, {Breytenbach}, {Buddelmeijer}, {Burke},
  {Calderone}, {Cano Rodr{\'\i}guez}, {Cara}, {Cardoso}, {Cheedella}, {Copin},
  {Corrales}, {Crichton}, {D'Avella}, {Deil}, {Depagne}, {Dietrich}, {Donath},
  {Droettboom}, {Earl}, {Erben}, {Fabbro}, {Ferreira}, {Finethy}, {Fox},
  {Garrison}, {Gibbons}, {Goldstein}, {Gommers}, {Greco}, {Greenfield},
  {Groener}, {Grollier}, {Hagen}, {Hirst}, {Homeier}, {Horton}, {Hosseinzadeh},
  {Hu}, {Hunkeler}, {Ivezi{\'c}}, {Jain}, {Jenness}, {Kanarek}, {Kendrew},
  {Kern}, {Kerzendorf}, {Khvalko}, {King}, {Kirkby}, {Kulkarni}, {Kumar},
  {Lee}, {Lenz}, {Littlefair}, {Ma}, {Macleod}, {Mastropietro}, {McCully},
  {Montagnac}, {Morris}, {Mueller}, {Mumford}, {Muna}, {Murphy}, {Nelson},
  {Nguyen}, {Ninan}, {N{\"o}the}, {Ogaz}, {Oh}, {Parejko}, {Parley}, {Pascual},
  {Patil}, {Patil}, {Plunkett}, {Prochaska}, {Rastogi}, {Reddy Janga},
  {Sabater}, {Sakurikar}, {Seifert}, {Sherbert}, {Sherwood-Taylor}, {Shih},
  {Sick}, {Silbiger}, {Singanamalla}, {Singer}, {Sladen}, {Sooley},
  {Sornarajah}, {Streicher}, {Teuben}, {Thomas}, {Tremblay}, {Turner},
  {Terr{\'o}n}, {van Kerkwijk}, {de la Vega}, {Watkins}, {Weaver}, {Whitmore},
  {Woillez}, {Zabalza}, \& {Astropy Contributors}}]{Astropy2018}
{Astropy Collaboration}, {Price-Whelan}, A.~M., {Sip{\H{o}}cz}, B.~M., {et~al.}
  2018, \aj, 156, 123

\bibitem[{{Astropy Collaboration} {et~al.}(2013){Astropy Collaboration},
  {Robitaille}, {Tollerud}, {Greenfield}, {Droettboom}, {Bray}, {Aldcroft},
  {Davis}, {Ginsburg}, {Price-Whelan}, {Kerzendorf}, {Conley}, {Crighton},
  {Barbary}, {Muna}, {Ferguson}, {Grollier}, {Parikh}, {Nair}, {Unther},
  {Deil}, {Woillez}, {Conseil}, {Kramer}, {Turner}, {Singer}, {Fox}, {Weaver},
  {Zabalza}, {Edwards}, {Azalee Bostroem}, {Burke}, {Casey}, {Crawford},
  {Dencheva}, {Ely}, {Jenness}, {Labrie}, {Lim}, {Pierfederici}, {Pontzen},
  {Ptak}, {Refsdal}, {Servillat}, \& {Streicher}}]{Astropy2013}
{Astropy Collaboration}, {Robitaille}, T.~P., {Tollerud}, E.~J., {et~al.} 2013,
  \aap, 558, A33

\bibitem[{{Bally} {et~al.}(1987){Bally}, {Langer}, {Stark}, \&
  {Wilson}}]{Bally1987}
{Bally}, J., {Langer}, W.~D., {Stark}, A.~A., \& {Wilson}, R.~W. 1987, \apjl,
  312, L45

\bibitem[{{Barnes} {et~al.}(2020){Barnes}, {Kauffmann}, {Bigiel}, {Brinkmann},
  {Colombo}, {Guzm{\'a}n}, {Kim}, {Sz{\H{u}}cs}, {Wakelam}, {Aalto},
  {Albertsson}, {Evans}, {Glover}, {Goldsmith}, {Kramer}, {Menten},
  {Nishimura}, {Viti}, {Watanabe}, {Weiss}, {Wienen}, {Wiesemeyer}, \&
  {Wyrowski}}]{Barnes2020}
{Barnes}, A.~T., {Kauffmann}, J., {Bigiel}, F., {et~al.} 2020, \mnras, 497,
  1972

\bibitem[{{Bergin} {et~al.}(1999){Bergin}, {Plume}, {Williams}, \&
  {Myers}}]{Bergin1999}
{Bergin}, E.~A., {Plume}, R., {Williams}, J.~P., \& {Myers}, P.~C. 1999, \apj,
  512, 724

\bibitem[{{Bern{\'e}} {et~al.}(2014){Bern{\'e}}, {Marcelino}, \&
  {Cernicharo}}]{Berne2014}
{Bern{\'e}}, O., {Marcelino}, N., \& {Cernicharo}, J. 2014, \apj, 795, 13

\bibitem[{{Black} \& {van Dishoeck}(1991)}]{Black1991}
{Black}, J.~H. \& {van Dishoeck}, E.~F. 1991, \apjl, 369, L9

\bibitem[{{Bron} {et~al.}(2021){Bron}, {Roueff}, {Gerin}, {Pety}, {Gratier},
  {Le Petit}, {Guzman}, {Orkisz}, {de Souza Magalhaes}, {Gaudel}, {Vono},
  {Bardeau}, {Chainais}, {Goicoechea}, {Hughes}, {Kainulainen}, {Languignon},
  {Le Bourlot}, {Levrier}, {Liszt}, {{\"O}berg}, {Peretto}, {Roueff}, \&
  {Sievers}}]{Bron2021}
{Bron}, E., {Roueff}, E., {Gerin}, M., {et~al.} 2021, \aap, 645, A28

\bibitem[{Burnham \& Anderson(2004)}]{Burnham2004}
Burnham, K.~P. \& Anderson, D.~R. 2004, Sociological methods \& research, 33,
  261

\bibitem[{{Caselli}(2002)}]{Caselli2002}
{Caselli}, P. 2002, \planss, 50, 1133

\bibitem[{{Caselli} {et~al.}(1998){Caselli}, {Walmsley}, {Terzieva}, \&
  {Herbst}}]{Caselli1998}
{Caselli}, P., {Walmsley}, C.~M., {Terzieva}, R., \& {Herbst}, E. 1998, \apj,
  499, 234

\bibitem[{{Cooksy} {et~al.}(1986){Cooksy}, {Blake}, \& {Saykally}}]{Cooksy1986}
{Cooksy}, A.~L., {Blake}, G.~A., \& {Saykally}, R.~J. 1986, \apjl, 305, L89

\bibitem[{{Cuadrado} {et~al.}(2019){Cuadrado}, {Salas}, {Goicoechea},
  {Cernicharo}, {Tielens}, \& {B{\'a}ez-Rubio}}]{Cuadrado2019}
{Cuadrado}, S., {Salas}, P., {Goicoechea}, J.~R., {et~al.} 2019, \aap, 625, L3

\bibitem[{{Dickinson} {et~al.}(1977){Dickinson}, {Phillips}, {Goldsmith},
  {Percival}, \& {Richards}}]{Dickinson1977}
{Dickinson}, A.~S., {Phillips}, T.~G., {Goldsmith}, P.~F., {Percival}, I.~C.,
  \& {Richards}, D. 1977, \aap, 54, 645

\bibitem[{{Dumouchel} {et~al.}(2010){Dumouchel}, {Faure}, \&
  {Lique}}]{Dumouchel2010}
{Dumouchel}, F., {Faure}, A., \& {Lique}, F. 2010, \mnras, 406, 2488

\bibitem[{{Dupree}(1974)}]{Dupree1974}
{Dupree}, A.~K. 1974, \apj, 187, 25

\bibitem[{{Dutrey} {et~al.}(1993){Dutrey}, {Duvert}, {Castets}, {Langer},
  {Bally}, \& {Wilson}}]{Dutrey1993}
{Dutrey}, A., {Duvert}, G., {Castets}, A., {et~al.} 1993, \aap, 270, 468

\bibitem[{{Evans}(1999)}]{Evans1999}
{Evans}, Neal~J., I. 1999, \araa, 37, 311

\bibitem[{{Faure} {et~al.}(2007){Faure}, {Varambhia}, {Stoecklin}, \&
  {Tennyson}}]{Faure2007}
{Faure}, A., {Varambhia}, H.~N., {Stoecklin}, T., \& {Tennyson}, J. 2007,
  \mnras, 382, 840

\bibitem[{{Friesen} {et~al.}(2017){Friesen}, {Pineda}, {co-PIs}, {Rosolowsky},
  {Alves}, {Chac{\'o}n-Tanarro}, {How-Huan Chen}, {Chun-Yuan Chen}, {Di
  Francesco}, {Keown}, {Kirk}, {Punanova}, {Seo}, {Shirley}, {Ginsburg},
  {Hall}, {Offner}, {Singh}, {Arce}, {Caselli}, {Goodman}, {Martin}, {Matzner},
  {Myers}, {Redaelli}, \& {GAS Collaboration}}]{Friesen2017}
{Friesen}, R.~K., {Pineda}, J.~E., {co-PIs}, {et~al.} 2017, \apj, 843, 63

\bibitem[{{Gao} \& {Solomon}(2004)}]{Gao2004}
{Gao}, Y. \& {Solomon}, P.~M. 2004, \apj, 606, 271

\bibitem[{{Gatley} {et~al.}(1974){Gatley}, {Becklin}, {Mattews}, {Neugebauer},
  {Penston}, \& {Scoville}}]{Gatley1974}
{Gatley}, I., {Becklin}, E.~E., {Mattews}, K., {et~al.} 1974, \apjl, 191, L121

\bibitem[{{Ginsburg} \& {Mirocha}(2011)}]{Ginsburg2011}
{Ginsburg}, A. \& {Mirocha}, J. 2011, {PySpecKit: Python Spectroscopic Toolkit}

\bibitem[{{Goicoechea} {et~al.}(2009){Goicoechea}, {Pety}, {Gerin},
  {Hily-Blant}, \& {Le Bourlot}}]{Goicoechea2009}
{Goicoechea}, J.~R., {Pety}, J., {Gerin}, M., {Hily-Blant}, P., \& {Le
  Bourlot}, J. 2009, \aap, 498, 771

\bibitem[{{Goicoechea} {et~al.}(2015){Goicoechea}, {Teyssier}, {Etxaluze},
  {Goldsmith}, {Ossenkopf}, {Gerin}, {Bergin}, {Black}, {Cernicharo},
  {Cuadrado}, {Encrenaz}, {Falgarone}, {Fuente}, {Hacar}, {Lis}, {Marcelino},
  {Melnick}, {M{\"u}ller}, {Persson}, {Pety}, {R{\"o}llig}, {Schilke}, {Simon},
  {Snell}, \& {Stutzki}}]{Goicoechea2015}
{Goicoechea}, J.~R., {Teyssier}, D., {Etxaluze}, M., {et~al.} 2015, \apj, 812,
  75

\bibitem[{{Goldsmith} \& {Kauffmann}(2017)}]{Goldsmith2017}
{Goldsmith}, P.~F. \& {Kauffmann}, J. 2017, \apj, 841, 25

\bibitem[{{Goldsmith} {et~al.}(2012){Goldsmith}, {Langer}, {Pineda}, \&
  {Velusamy}}]{Goldsmith2012}
{Goldsmith}, P.~F., {Langer}, W.~D., {Pineda}, J.~L., \& {Velusamy}, T. 2012,
  \apjs, 203, 13

\bibitem[{{Gordon} \& {Sorochenko}(2009)}]{Gordon2009}
{Gordon}, M.~A. \& {Sorochenko}, R.~L., eds. 2009, Astrophysics and Space
  Science Library, Vol. 282, {Radio Recombination Lines}

\bibitem[{{Guevara} {et~al.}(2020){Guevara}, {Stutzki}, {Ossenkopf-Okada},
  {Simon}, {P{\'e}rez-Beaupuits}, {Beuther}, {Bihr}, {Higgins}, {Graf}, \&
  {G{\"u}sten}}]{Guevara2020}
{Guevara}, C., {Stutzki}, J., {Ossenkopf-Okada}, V., {et~al.} 2020, \aap, 636,
  A16

\bibitem[{{Habing}(1968)}]{Habing1968}
{Habing}, H.~J. 1968, \bain, 19, 421

\bibitem[{{Hacar} {et~al.}(2020){Hacar}, {Bosman}, \& {van
  Dishoeck}}]{Hacar2020}
{Hacar}, A., {Bosman}, A.~D., \& {van Dishoeck}, E.~F. 2020, \aap, 635, A4

\bibitem[{{Hacar} {et~al.}(2018){Hacar}, {Tafalla}, {Forbrich}, {Alves},
  {Meingast}, {Grossschedl}, \& {Teixeira}}]{Hacar2018}
{Hacar}, A., {Tafalla}, M., {Forbrich}, J., {et~al.} 2018, \aap, 610, A77

\bibitem[{{Herbst} \& {Klemperer}(1973)}]{Herbst1973}
{Herbst}, E. \& {Klemperer}, W. 1973, \apj, 185, 505

\bibitem[{{Heyminck} {et~al.}(2012){Heyminck}, {Graf}, {G{\"u}sten}, {Stutzki},
  {H{\"u}bers}, \& {Hartogh}}]{Heyminck2012}
{Heyminck}, S., {Graf}, U.~U., {G{\"u}sten}, R., {et~al.} 2012, \aap, 542, L1

\bibitem[{{Hoang-Binh} \& {Walmsley}(1974)}]{Hoang-Bihn1974}
{Hoang-Binh}, D. \& {Walmsley}, C.~M. 1974, \aap, 35, 49

\bibitem[{{Hollenbach} {et~al.}(1991){Hollenbach}, {Takahashi}, \&
  {Tielens}}]{Hollenbach1991}
{Hollenbach}, D.~J., {Takahashi}, T., \& {Tielens}, A.~G.~G.~M. 1991, \apj,
  377, 192

\bibitem[{{Hollenbach} \& {Tielens}(1999)}]{Hollenbach1999}
{Hollenbach}, D.~J. \& {Tielens}, A.~G.~G.~M. 1999, Reviews of Modern Physics,
  71, 173

\bibitem[{{Hunter}(2007)}]{Hunter2007}
{Hunter}, J.~D. 2007, Computing in Science and Engineering, 9, 90

\bibitem[{{Kauffmann} {et~al.}(2017){Kauffmann}, {Goldsmith}, {Melnick},
  {Tolls}, {Guzman}, \& {Menten}}]{Kauffmann2017}
{Kauffmann}, J., {Goldsmith}, P.~F., {Melnick}, G., {et~al.} 2017, \aap, 605,
  L5

\bibitem[{{Kaufman} {et~al.}(2006){Kaufman}, {Wolfire}, \&
  {Hollenbach}}]{Kaufman2006}
{Kaufman}, M.~J., {Wolfire}, M.~G., \& {Hollenbach}, D.~J. 2006, \apj, 644, 283

\bibitem[{{Klein} {et~al.}(2012){Klein}, {Hochg{\"u}rtel}, {Kr{\"a}mer},
  {Bell}, {Meyer}, \& {G{\"u}sten}}]{KleinHochgurtel:2012aa}
{Klein}, B., {Hochg{\"u}rtel}, S., {Kr{\"a}mer}, I., {et~al.} 2012, \aap, 542,
  L3

\bibitem[{{Kraus}(2009)}]{Kraus:2009aa}
{Kraus}, A. 2009, Calibration of the Effelsberg 100m telescope, Tech. rep.,
  Max-Planck-Institut f\"ur Radioastronomie

\bibitem[{{Kritsuk} \& {Norman}(2004)}]{Kritsuk2004}
{Kritsuk}, A.~G. \& {Norman}, M.~L. 2004, \apjl, 601, L55

\bibitem[{{Kritsuk} {et~al.}(2017){Kritsuk}, {Ustyugov}, \&
  {Norman}}]{Kritsuk2017}
{Kritsuk}, A.~G., {Ustyugov}, S.~D., \& {Norman}, M.~L. 2017, New Journal of
  Physics, 19, 065003

\bibitem[{{Krumholz}(2014)}]{Krumholz2014}
{Krumholz}, M.~R. 2014, \mnras, 437, 1662

\bibitem[{{Kutner} {et~al.}(1985){Kutner}, {Machnik}, {Mead}, \&
  {Evans}}]{Kutner1985}
{Kutner}, M.~L., {Machnik}, D.~E., {Mead}, K.~N., \& {Evans}, N.~J., I. 1985,
  \apj, 299, 351

\bibitem[{{Langer} \& {Penzias}(1990)}]{Langer1990}
{Langer}, W.~D. \& {Penzias}, A.~A. 1990, \apj, 357, 477

\bibitem[{{Lenz} \& {Ayres}(1992)}]{Lenz1992}
{Lenz}, D.~D. \& {Ayres}, T.~R. 1992, \pasp, 104, 1104

\bibitem[{{Lombardi} {et~al.}(2014){Lombardi}, {Bouy}, {Alves}, \&
  {Lada}}]{Lombardi2014}
{Lombardi}, M., {Bouy}, H., {Alves}, J., \& {Lada}, C.~J. 2014, \aap, 566, A45

\bibitem[{{Luisi} {et~al.}(2019){Luisi}, {Anderson}, {Liu}, {Anish Roshi}, \&
  {Churchwell}}]{Luisi2019}
{Luisi}, M., {Anderson}, L.~D., {Liu}, B., {Anish Roshi}, D., \& {Churchwell},
  E. 2019, \apjs, 241, 2

\bibitem[{{McKee}(1989)}]{McKee1989}
{McKee}, C.~F. 1989, \apj, 345, 782

\bibitem[{{McKee} \& {Ostriker}(2007)}]{McKee2007}
{McKee}, C.~F. \& {Ostriker}, E.~C. 2007, \araa, 45, 565

\bibitem[{{Megeath} {et~al.}(2012){Megeath}, {Gutermuth}, {Muzerolle},
  {Kryukova}, {Flaherty}, {Hora}, {Allen}, {Hartmann}, {Myers}, {Pipher},
  {Stauffer}, {Young}, \& {Fazio}}]{Megeath2012}
{Megeath}, S.~T., {Gutermuth}, R., {Muzerolle}, J., {et~al.} 2012, \aj, 144,
  192

\bibitem[{{Meijerink} {et~al.}(2011){Meijerink}, {Spaans}, {Loenen}, \& {van
  der Werf}}]{Meijerink2011}
{Meijerink}, R., {Spaans}, M., {Loenen}, A.~F., \& {van der Werf}, P.~P. 2011,
  \aap, 525, A119

\bibitem[{{Melnick} {et~al.}(2011){Melnick}, {Tolls}, {Snell}, {Bergin},
  {Hollenbach}, {Kaufman}, {Li}, \& {Neufeld}}]{Melnick2011}
{Melnick}, G.~J., {Tolls}, V., {Snell}, R.~L., {et~al.} 2011, \apj, 727, 13

\bibitem[{{Menten} {et~al.}(2007){Menten}, {Reid}, {Forbrich}, \&
  {Brunthaler}}]{Menten2007}
{Menten}, K.~M., {Reid}, M.~J., {Forbrich}, J., \& {Brunthaler}, A. 2007, \aap,
  474, 515

\bibitem[{{Menzel}(1968)}]{Menzel1968}
{Menzel}, D.~H. 1968, \nat, 218, 756

\bibitem[{{Mestel} \& {Spitzer}(1956)}]{Mestel1956}
{Mestel}, L. \& {Spitzer}, L., J. 1956, \mnras, 116, 503

\bibitem[{{Morris} {et~al.}(1974){Morris}, {Zuckerman}, {Turner}, \&
  {Palmer}}]{Morris1974}
{Morris}, M., {Zuckerman}, B., {Turner}, B.~E., \& {Palmer}, P. 1974, \apjl,
  192, L27

\bibitem[{{Natta} {et~al.}(1994){Natta}, {Walmsley}, \& {Tielens}}]{Natta1994}
{Natta}, A., {Walmsley}, C.~M., \& {Tielens}, A.~G.~G.~M. 1994, \apj, 428, 209

\bibitem[{{Neufeld} \& {Wolfire}(2017)}]{Neufeld2017}
{Neufeld}, D.~A. \& {Wolfire}, M.~G. 2017, \apj, 845, 163

\bibitem[{Newville {et~al.}(2014)Newville, Stensitzki, Allen, \&
  Ingargiola}]{newville_2014_11813}
Newville, M., Stensitzki, T., Allen, D.~B., \& Ingargiola, A. 2014, {LMFIT:
  Non-Linear Least-Square Minimization and Curve-Fitting for Python}

\bibitem[{{Nielbock} {et~al.}(2003){Nielbock}, {Chini}, \&
  {M{\"u}ller}}]{Nielbock2003}
{Nielbock}, M., {Chini}, R., \& {M{\"u}ller}, S.~A.~H. 2003, \aap, 408, 245

\bibitem[{{Ohashi} {et~al.}(2014){Ohashi}, {Tatematsu}, {Choi}, {Kang},
  {Umemoto}, {Lee}, {Hirota}, {Yamamoto}, \& {Mizuno}}]{Ohashi2014}
{Ohashi}, S., {Tatematsu}, K., {Choi}, M., {et~al.} 2014, \pasj, 66, 119

\bibitem[{{Oppenheimer} \& {Dalgarno}(1974)}]{Oppenheimer1974}
{Oppenheimer}, M. \& {Dalgarno}, A. 1974, \apj, 192, 29

\bibitem[{{Ossenkopf} {et~al.}(2013){Ossenkopf}, {R{\"o}llig}, {Neufeld},
  {Pilleri}, {Lis}, {Fuente}, {van der Tak}, \& {Bergin}}]{Ossenkopf2013}
{Ossenkopf}, V., {R{\"o}llig}, M., {Neufeld}, D.~A., {et~al.} 2013, \aap, 550,
  A57

\bibitem[{{Ott} {et~al.}(1994){Ott}, {Witzel}, {Quirrenbach}, {Krichbaum},
  {Standke}, {Schalinski}, \& {Hummel}}]{Ott1994}
{Ott}, M., {Witzel}, A., {Quirrenbach}, A., {et~al.} 1994, \aap, 284, 331

\bibitem[{Pabst {et~al.}(2019)Pabst, Higgins, Goicoechea, Teyssier, Berne,
  Chambers, Wolfire, Suri, Guesten, Stutzki, Graf, Risacher, \&
  Tielens}]{Pabst2019}
Pabst, C., Higgins, R., Goicoechea, J.~R., {et~al.} 2019, Nature

\bibitem[{{Pabst} {et~al.}(2020){Pabst}, {Goicoechea}, {Teyssier}, {Bern{\'e}},
  {Higgins}, {Chambers}, {Kabanovic}, {G{\"u}sten}, {Stutzki}, \&
  {Tielens}}]{Pabst2020}
{Pabst}, C.~H.~M., {Goicoechea}, J.~R., {Teyssier}, D., {et~al.} 2020, \aap,
  639, A2

\bibitem[{{Pety} {et~al.}(2017){Pety}, {Guzm{\'a}n}, {Orkisz}, {Liszt},
  {Gerin}, {Bron}, {Bardeau}, {Goicoechea}, {Gratier}, {Le Petit}, {Levrier},
  {{\"O}berg}, {Roueff}, \& {Sievers}}]{Pety2017}
{Pety}, J., {Guzm{\'a}n}, V.~V., {Orkisz}, J.~H., {et~al.} 2017, \aap, 599, A98

\bibitem[{{Pound} \& {Wolfire}(2008)}]{Pound2008}
{Pound}, M.~W. \& {Wolfire}, M.~G. 2008, in Astronomical Society of the Pacific
  Conference Series, Vol. 394, Astronomical Data Analysis Software and Systems
  XVII, ed. R.~W. {Argyle}, P.~S. {Bunclark}, \& J.~R. {Lewis}, 654

\bibitem[{{Reid} {et~al.}(2009){Reid}, {Menten}, {Zheng}, {Brunthaler},
  {Moscadelli}, {Xu}, {Zhang}, {Sato}, {Honma}, {Hirota}, {Hachisuka}, {Choi},
  {Moellenbrock}, \& {Bartkiewicz}}]{Reid2009}
{Reid}, M.~J., {Menten}, K.~M., {Zheng}, X.~W., {et~al.} 2009, \apj, 700, 137

\bibitem[{{Reipurth} {et~al.}(1999){Reipurth}, {Rodr{\'\i}guez}, \&
  {Chini}}]{Reipurth1999}
{Reipurth}, B., {Rodr{\'\i}guez}, L.~F., \& {Chini}, R. 1999, \aj, 118, 983

\bibitem[{{Risacher} {et~al.}(2016){Risacher}, {G{\"u}sten}, {Stutzki},
  {H{\"u}bers}, {Bell}, {Buchbender}, {B{\"u}chel}, {Csengeri}, {Graf},
  {Heyminck}, {Higgins}, {Honingh}, {Jacobs}, {Klein}, {Okada}, {Parikka},
  {P{\"u}tz}, {Reyes}, {Ricken}, {Riquelme}, {Simon}, \&
  {Wiesemeyer}}]{Risacher2016}
{Risacher}, C., {G{\"u}sten}, R., {Stutzki}, J., {et~al.} 2016, \aap, 595, A34

\bibitem[{{Roueff} {et~al.}(2015){Roueff}, {Loison}, \& {Hickson}}]{Roueff2015}
{Roueff}, E., {Loison}, J.~C., \& {Hickson}, K.~M. 2015, \aap, 576, A99

\bibitem[{{Roy} {et~al.}(2004){Roy}, {Teuber}, \& {Keller}}]{Roy2004}
{Roy}, A.~L., {Teuber}, U., \& {Keller}, R. 2004, in European VLBI Network on
  New Developments in VLBI Science and Technology, 265--270

\bibitem[{Salas {et~al.}(2016)Salas, Morabito, Salgado, Oonk, \&
  Tielens}]{crrlpy}
Salas, P., Morabito, L., Salgado, F., Oonk, R., \& Tielens, A. 2016, CRRLpy:
  First pre-release

\bibitem[{{Salas} {et~al.}(2019){Salas}, {Oonk}, {Emig}, {Pabst}, {Toribio},
  {R{\"o}ttgering}, \& {Tielens}}]{Salas2019}
{Salas}, P., {Oonk}, J.~B.~R., {Emig}, K.~L., {et~al.} 2019, \aap, 626, A70

\bibitem[{{Salgado} {et~al.}(2017{\natexlab{a}}){Salgado}, {Morabito}, {Oonk},
  {Salas}, {Toribio}, {R{\"o}ttgering}, \& {Tielens}}]{Salgado2017a}
{Salgado}, F., {Morabito}, L.~K., {Oonk}, J.~B.~R., {et~al.}
  2017{\natexlab{a}}, \apj, 837, 141

\bibitem[{{Salgado} {et~al.}(2017{\natexlab{b}}){Salgado}, {Morabito}, {Oonk},
  {Salas}, {Toribio}, {R{\"o}ttgering}, \& {Tielens}}]{Salgado2017b}
{Salgado}, F., {Morabito}, L.~K., {Oonk}, J.~B.~R., {et~al.}
  2017{\natexlab{b}}, \apj, 837, 142

\bibitem[{{Shirley}(2015)}]{Shirley2015}
{Shirley}, Y.~L. 2015, \pasp, 127, 299

\bibitem[{{Siddiqui} {et~al.}(2020){Siddiqui}, {Khan}, \&
  {Qaiyum}}]{Siddiqui2020}
{Siddiqui}, S.~A., {Khan}, S., \& {Qaiyum}, A. 2020, \mnras, 492, 1049

\bibitem[{{Sofia} {et~al.}(2004){Sofia}, {Lauroesch}, {Meyer}, \&
  {Cartledge}}]{Sofia2004}
{Sofia}, U.~J., {Lauroesch}, J.~T., {Meyer}, D.~M., \& {Cartledge}, S.~I.~B.
  2004, \apj, 605, 272

\bibitem[{{Sorochenko} \& {Tsivilev}(2010)}]{Sorochenko2010}
{Sorochenko}, R.~L. \& {Tsivilev}, A.~P. 2010, Kinematics and Physics of
  Celestial Bodies, 26, 162

\bibitem[{{Tatematsu} {et~al.}(1993){Tatematsu}, {Umemoto}, {Kameya}, {Hirano},
  {Hasegawa}, {Hayashi}, {Iwata}, {Kaifu}, {Mikami}, {Murata}, {Nakano},
  {Nakano}, {Ohashi}, {Sunada}, {Takaba}, \& {Yamamoto}}]{Tatematsu1993}
{Tatematsu}, K., {Umemoto}, T., {Kameya}, O., {et~al.} 1993, \apj, 404, 643

\bibitem[{{Tielens}(2005)}]{Tielens2005}
{Tielens}, A.~G.~G.~M. 2005, {The Physics and Chemistry of the Interstellar
  Medium}

\bibitem[{{Tsivilev}(2014)}]{Tsivilev2014}
{Tsivilev}, A.~P. 2014, Astronomy Letters, 40, 615

\bibitem[{{van der Tak} {et~al.}(2007){van der Tak}, {Black}, {Sch{\"o}ier},
  {Jansen}, \& {van Dishoeck}}]{vanderTak2007}
{van der Tak}, F.~F.~S., {Black}, J.~H., {Sch{\"o}ier}, F.~L., {Jansen}, D.~J.,
  \& {van Dishoeck}, E.~F. 2007, \aap, 468, 627

\bibitem[{{van der Tak} \& {van Dishoeck}(2000)}]{vanderTak2000}
{van der Tak}, F.~F.~S. \& {van Dishoeck}, E.~F. 2000, \aap, 358, L79

\bibitem[{{van der Wiel} {et~al.}(2009){van der Wiel}, {van der Tak},
  {Ossenkopf}, {Spaans}, {Roberts}, {Fuller}, \& {Plume}}]{vanderWiel2009}
{van der Wiel}, M.~H.~D., {van der Tak}, F.~F.~S., {Ossenkopf}, V., {et~al.}
  2009, \aap, 498, 161

\bibitem[{{Vollmer} {et~al.}(2017){Vollmer}, {Gratier}, {Braine}, \&
  {Bot}}]{Vollmer2017}
{Vollmer}, B., {Gratier}, P., {Braine}, J., \& {Bot}, C. 2017, \aap, 602, A51

\bibitem[{{Walmsley} \& {Watson}(1982)}]{Walmsley1982}
{Walmsley}, C.~M. \& {Watson}, W.~D. 1982, \apj, 260, 317

\bibitem[{{Watson}(1972)}]{Watson1972}
{Watson}, W.~D. 1972, \apj, 176, 103

\bibitem[{{Wiese} \& {Fuhr}(2007)}]{Wiese2007}
{Wiese}, W.~L. \& {Fuhr}, J.~R. 2007, Journal of Physical and Chemical
  Reference Data, 36, 1287

\bibitem[{{Winkel} {et~al.}(2012){Winkel}, {Kraus}, \&
  {Bach}}]{WinkelKraus:2012ab}
{Winkel}, B., {Kraus}, A., \& {Bach}, U. 2012, \aap, 540, A140

\bibitem[{{Wyrowski} {et~al.}(1997){Wyrowski}, {Schilke}, {Hofner}, \&
  {Walmsley}}]{Wyrowski1997}
{Wyrowski}, F., {Schilke}, P., {Hofner}, P., \& {Walmsley}, C.~M. 1997, \apjl,
  487, L171

\bibitem[{{Wyrowski} {et~al.}(2000){Wyrowski}, {Walmsley}, {Goss}, \&
  {Tielens}}]{Wyrowski2000}
{Wyrowski}, F., {Walmsley}, C.~M., {Goss}, W.~M., \& {Tielens}, A.~G.~G.~M.
  2000, \apj, 543, 245

\bibitem[{{Young} {et~al.}(2012){Young}, {Becklin}, {Marcum}, {Roellig}, {De
  Buizer}, {Herter}, {G{\"u}sten}, {Dunham}, {Temi}, {Andersson}, {Backman},
  {Burgdorf}, {Caroff}, {Casey}, {Davidson}, {Erickson}, {Gehrz}, {Harper},
  {Harvey}, {Helton}, {Horner}, {Howard}, {Klein}, {Krabbe}, {McLean}, {Meyer},
  {Miles}, {Morris}, {Reach}, {Rho}, {Richter}, {Roeser}, {Sandell}, {Sankrit},
  {Savage}, {Smith}, {Shuping}, {Vacca}, {Vaillancourt}, {Wolf}, \&
  {Zinnecker}}]{Young2012}
{Young}, E.~T., {Becklin}, E.~E., {Marcum}, P.~M., {et~al.} 2012, \apjl, 749,
  L17

\bibitem[{{Zari} {et~al.}(2017){Zari}, {Brown}, {de Bruijne}, {Manara}, \& {de
  Zeeuw}}]{Zari2017}
{Zari}, E., {Brown}, A.~G.~A., {de Bruijne}, J., {Manara}, C.~F., \& {de
  Zeeuw}, P.~T. 2017, \aap, 608, A148

\bibitem[{{Zhang} {et~al.}(2020){Zhang}, {Ren}, {Wu}, {Li}, {Zhu}, {Zhang},
  {Mardones}, {Wang}, {Shi}, {Yue}, {Luo}, {Xie}, {Jiao}, {Liu}, {Xu}, \&
  {Wang}}]{Zhang2020}
{Zhang}, C., {Ren}, Z., {Wu}, J., {et~al.} 2020, \mnras, 497, 793

\end{thebibliography}

% \appendix
\begin{appendix}

\section{Best fit Gaussian components}

Here we provide the best fit line properties for the [CII] line.

% \begin{minipage}[t]{\textwidth}
\begin{table}[!htb]
% \begin{center}
% \begin{longtable}{lcccccc}
\caption{\label{tab:cii} Best fit [CII] line properties.}
% \centering
\begin{tabular}{lcccccc}
\toprule
Region & Line ID &  $T_{a}$     &  $\varv_{\mathrm{c}}$ &  $\Delta\varv$\tablefootmark{a}  & $\int T_{\mathrm{a}}^{*}dv$         & rms  \\%& SNR\tablefootmark{b} \\
       &         & (K)          & (km s$^{-1}$)         & (km s$^{-1}$)                    & (K km s$^{-1}$)                     & (K)  \\%&     \\
\midrule
  \multirow{3}{*}{EFF1} & a & $27\pm1$     & $9.90\pm0.08$ & $3.4\pm0.1$    & $97\pm5$     &  $0.35$ \\%& $227$ \\
                        & b & $4.9\pm1.3$  & $12.6\pm0.4$  & $3.3\pm0.5$    & $16.9\pm4.9$ &  $0.35$ \\%& $40$  \\
                        & c & $0.8\pm0.1$  & $5.9\pm0.4$   & $2.3\pm0.7$    & $2.2\pm0.8$  &  $0.35$ \\%& $6$   \\
  \hline
  \multirow{4}{*}{EFF2} & a & $16.7\pm2.8$ & $10.96\pm0.05$ & $2.8\pm0.2$   & $50.83\pm9.08$ &  $0.33$ \\%&  $134$ \\
                        & b & $3.9\pm0.3$  & $11.62\pm0.03$ & $0.82\pm0.07$ & $3.4\pm0.4$    &  $0.33$ \\%&   $17$ \\
                        & c & $2.9\pm0.8$  & $7.87\pm0.07$  & $1.6\pm0.2$   & $5.4\pm1.6$    &  $0.33$ \\%&   $18$ \\
                        & d & $6.6\pm2.9$  & $10.6\pm0.1$   & $5.4\pm0.5$   & $38.6\pm17.4$  &  $0.33$ \\%&   $74$ \\
  \hline
  \multirow{3}{*}{EFF3} & a & $10.7\pm0.7$ & $10.09\pm0.05$ & $1.34\pm0.07$ & $15.3\pm1.4$ &  $0.35$ \\%& $56$ \\
                        & b & $5.7\pm0.6$  & $11.07\pm0.06$ & $4.7\pm0.2$   & $29.7\pm3.5$ &  $0.35$ \\%& $58$ \\
                        & c & $10.3\pm0.7$ & $11.4\pm0.07$  & $1.5\pm0.1$   & $17.0\pm1.7$ &  $0.35$ \\%& $58$ \\
  \hline
  \multirow{3}{*}{EFF4} & a & $23.5\pm1.0$ & $9.81\pm0.02$  & $1.84\pm0.05$ & $46.9\pm2.3$ &  $0.36$ \\%& $141$ \\
                        & b & $16.9\pm1.3$ & $11.78\pm0.03$ & $2.05\pm0.09$ & $37.8\pm3.4$ &  $0.36$ \\%& $107$ \\
                        & c & $6.4\pm1.6$  & $11.02\pm0.08$ & $4.7\pm0.2$   & $33.4\pm8.9$ &  $0.36$ \\%& $63$  \\
  \hline
  \multirow{3}{*}{EFF4\tablefootmark{b}} & a\tablefootmark{c} & $64\pm15$   & $10.74\pm0.05$ & $2.6\pm0.1$ & $183\pm44$  &  $0.36$ \\%& $465$ \\
                                         & b\tablefootmark{d} & $40\pm14$   & $10.86\pm0.03$ & $1.8\pm0.1$ & $80\pm29$   &  $0.36$ \\%& $243$ \\
                                         & c                  & $2.3\pm1.7$ & $13\pm2$       & $3.7\pm1.8$ & $9\pm8$     &  $0.36$ \\%& $20$  \\
                                         & d                  & $0.8\pm0.2$ & $7.2\pm0.2$    & $1.2\pm0.5$ & $1.0\pm0.5$ &  $0.36$ \\%& $3.8$  \\
  \hline
  \multirow{5}{*}{EFF5} & a & $38.2\pm3.7$   & $10.71\pm0.03$ & $1.81\pm0.05$ & $73.6\pm7.5$   &  $0.38$ \\%&  $217$ \\
                        & b & $26.32\pm5.02$ & $11.85\pm0.04$ & $3.8\pm0.1$   & $105.6\pm20.6$ &  $0.38$ \\%&  $216$ \\
                        & c & $ 9.1\pm2.5$   & $11.91\pm0.04$ & $0.89\pm0.09$ & $8.6\pm2.5$    &  $0.38$ \\%&   $36$ \\
                        & d & $ 8.3\pm2.4$   & $13.31\pm0.06$ & $1.0\pm0.1$   & $9.6\pm3.1$    &  $0.38$ \\%&   $37$ \\
                        & f & $ 0.5\pm0.1$   & $17.6\pm0.3$   & $1.4\pm0.7$   & $0.8\pm0.5$    &  $0.38$ \\%&   $2.7$ \\
  
  \bottomrule
\end{tabular}
\tablefoot{
\tablefoottext{a}{Full width at half maximum of the line, $\Delta\varv=2\sqrt{2\ln(2)}\sigma_{\varv}$ with $\sigma_{\varv}$ the standard deviation of the Gaussian profile.}\\
\tablefoottext{b}{Model considering self-absorption.}\\
\tablefoottext{c}{Background component.}\\
\tablefoottext{d}{Foreground component.}
}
% \end{center}
% \end{longtable}
\end{table}
% \end{minipage}

\end{appendix}

\end{document}